\documentclass{aa} 
\usepackage{graphicx}
\usepackage{amsmath,amssymb} 
\usepackage{txfonts}
\usepackage[T1]{fontenc}
\usepackage{afterpage}
\usepackage{amssymb}
\usepackage{natbib} 
\usepackage{wrapfig}
\usepackage{graphicx}
\usepackage{soul}
\usepackage{color, colortbl}
\usepackage{subfig}
\bibpunct{(}{)}{;}{a}{}{,}

 \def\lesssim{\mathrel{\hbox{\rlap{\hbox{\lower4pt\hbox{$\sim$}}}\hbox{$<$}}}}
 \def\gtrsim{\mathrel{\hbox{\rlap{\hbox{\lower4pt\hbox{$\sim$}}}\hbox{$>$}}}}

\begin{document}

   \title{The Traditional Approximation of Rotation\\ including the centrifugal acceleration for slightly deformed stars}

   \subtitle{}

   \author{S. Mathis\inst{1}, V. Prat\inst{1}}

   \institute{AIM, CEA, CNRS, Universit\'e Paris-Saclay, Universit\'e Paris Diderot, Sorbonne Paris Cit\'e, F-91191 Gif-sur-Yvette Cedex, France\\
              \email{stephane.mathis@cea.fr}
             }


 
  \abstract
   {The Traditional Approximation of Rotation (TAR) is a treatment of the dynamical equations of rotating stably stratified fluids where the action of the Coriolis acceleration along the direction of the entropy (and chemicals) stratification is neglected while assuming that the fluid motions are mostly horizontal because of their inhibition in the vertical direction by the buoyancy force. This leads to neglect the horizontal projection of the rotation vector in the equations for the dynamics of gravito-inertial waves (GIWs) that become separable as in the non-rotating case while they are not in the case where the full Coriolis acceleration is taken into account. This approximation, first introduced in geophysical fluid dynamics for thin atmospheres and oceans, has been broadly applied in stellar (and planetary) astrophysics to study low-frequency GIWs that have short vertical wavelengths. It is now tested thanks to direct 2D oscillation codes, which constrain its domain of validity. Its mathematical flexibility allows us to explore broad parameters space and to perform detailed seismic modelling of stars
.}
 {TAR is built on the assumptions that the star is spherical (i.e. its centrifugal deformation is neglected) and uniformly rotating while an adiabatic treatment of the dynamics of the waves is adopted, their induced gravitational potential fluctuations being neglected. However, it has been recently generalised with including the effects of a differential rotation. We aim to do a new generalisation that takes into account the centrifugal acceleration in the case of moderately uniformly rotating deformed stars.}
{We make an analytical expansion of the equations for the dynamics of GIWs in a spheroidal coordinates system with assuming the hierarchies of frequencies and amplitudes of the velocity components adopted within TAR in the spherical case.}
 {We derive the complete set of equations that generalizes TAR with taking into account the centrifugal acceleration. As in the case of a differentially rotating spherical star, the problem becomes 2D but can be treated analytically if assuming the anelastic and JWKB approximations, which are relevant for low-frequency GIWs. It allows us to derive a generalised Laplace tidal equation for the horizontal eigenfunctions and asymptotic wave periods that can be used to probe the structure and dynamics of rotating deformed stars thanks to asteroseismology. A first numerical exploration of its eigenvalues and horizontal eigenfunctions shows their variation as a function of the pseudo-radius for different rotation rates and frequencies and the development of avoided crossings.}
   {} 


   \maketitle
%

\section{Introduction}

The Traditional Approximation of Rotation (hereafter TAR) has first been introduced to study the dynamics of the shallow Earth atmosphere and oceans \citep[e.g. Eckart 1960,][Zeitlin 2018]{Gerkemaetal2008}. In particular, it is broadly used for the understanding of the propagation of waves in stably stratified (in entropy and chemical composition) and rotating atmospheric and oceanic layers. There, inertia-gravity waves, which are often called gravito-inertial waves in stellar physics \citep[e.g.][]{DintransRieutord1999}, propagate under the combined action of the buoyancy force and the Coriolis acceleration. Assuming that in the direction of the stable entropy/chermical stratification, the buoyancy force is stronger than the Coriolis acceleration (i.e. $2\Omega\ll N$, where $\Omega$ is the angular velocity and $N$ the Brunt-Va\"is\"al\"a frequency) leads to waves with horizontal velocities larger than vertical ones. This allows one to neglect the terms involving the latitudinal component of the rotation vector $\Omega_{\rm H}=\Omega\sin\theta$ (where $\theta$ is the colatitude) in the linearised hydrodynamical equations. In this framework, the (Poincar\'e) wave equation, which is bi-dimensional and non-separable in the general case, becomes separable \citep[e.g.][]{GerkemaShrira2005}. Therefore, scalar quantities (e.g. the density, pressure, temperature, and entropy waves' fluctuations) and the velocity components can be expressed as products of vertical/radial functions, special latitudinal functions: the so-called Hough functions that reduce to Legendre polynomials in the non-rotating case \citep{Hough1898,LonguetHiggins1968}, and Fourier series in time and azimuth.\\ 

In stellar physics, TAR and its flexibility has been extensively used to study low-frequency gravito-inertial waves (hereafter GIWs), including Rossby waves, both in single and in double stars \citep[e.g.][]{Berthomieuetal1978,LeeSaio1987,LeeSaio1989,LeeSaio1997,Townsend2003,Mathis2009}. It allows stellar physicists to derive powerful seismic diagnosis such that the period spacing between consecutive high radial orders gravito-inertial modes in uniformly and differentially rotating spherical stars \citep{Bouabidetal2013,Ouazzanietal2017,VanReethetal2018}. With the advent of space asteroseismology using high precision photometry \citep{AJCDK2010}, the period spacing derived within the TAR framework gives access to properties of the chemical stratification and to the rotation rate near the convective core of rapidly rotating intermediate-mass $\gamma$ Doradus stars for more than 60 stars \citep[][]{VanReethetal2015a,VanReethetal2015b,VanReethetal2016,Aertsetal2017,VanReethetal2018,Christopheetal2018,Lietal2019}. These quantities constitute key informations to improve our knowledge of stellar structure, evolution, internal angular momentum transport and to calibrate stellar models \citep[e.g.][and references therein]{Pedersenetal2018,Aertsetal2018a,Aertsetal2018b,Ouazzanietal2018}. However, in addition to assuming that $2\Omega\ll N$, other hypothesis are done to apply TAR in stellar interiors \citep[e.g.][]{LeeSaio1997,Townsend2003}. First, the rotation is assumed to be uniform. Next, we assume that the centrifugal distortion of the star can be neglected, i.e. $\Omega\ll\Omega_{\rm K}\equiv\sqrt{GM/R^3}$, where $G$, $M$, and $R$ are the universal constant of gravity, the mass of the star, and the stellar radius, respectively, and $\Omega_{\rm K}$ is the Keplerian critical angular velocity. The waves' fluctuation of the gravitational potential is neglected following \cite{Cowling1941}. Finally, the waves' motions are assumed to be adiabatic. 

Different ways can be considered to go beyond these assumptions. One can choose to use 2D oscillation codes \citep[][]{Reeseetal2006,Ballotetal2010,Ouazzanietal2012}. However, they are not available yet to the whole community while their needed computation time and resources can prevent detailed seismic modeling at this stage. One can alternatively build a 2D asymptotic theory for GIWs that go beyond the TAR \citep{Pratetal2016,Pratetal2018}. However, the associated derivation of the needed asymptotic seismic diagnosis is still in its infancy \citep{Pratetal2017} and should be developped. Finally, in the case of strongly stratified stellar radiation zones for which $2\Omega\ll N$, one can try to improve TAR by using the hierarchy between frequencies and the corresponding properties of motions. This has been done successfully to include the effects of general differential rotation on low-frequency GIWs \citep{OgilvieLin2004,Mathis2009}. In this case, the problem becomes 2D and non separable as in the general case where the full Coriolis acceleration is taken into account even in the case of a "shellular" rotation that depends on the radius only. However, the problem can be treated analytically in the case of low-frequency GIWs following \cite{OgilvieLin2004} and \cite{Mathis2009} who considered rapidly oscillating waves in the vertical direction with assuming the anelastic approximation where acoustic waves are filtered out. They introduced generalised 2D Hough functions that depend on the latitude and on the radius, the latter acting only as a parameter. The results of this method have been successfully applied in \cite{VanReethetal2018} to derive the variation of the asymptotic period spacing in the case of a weak radial differential rotation as observed in intermediate-mass stars using asteroseismology \citep[][]{Kurtzetal2014,Saioetal2015,Murphyetal2016,Aertsetal2017}.\\

In this theoretical work, we consider the case of "moderately" rapidly rotating stars (or planets). Their shape becomes a slightly deformed spheroid because of the action of the centrifugal acceleration. This case has been studied using 2D oscillation modes numerical computation in \cite{Ouazzanietal2017}. They demonstrated that the results obtained using TAR and the related assumptions (i.e. studying uniformly rotating spherical stars) are in qualitative agreement with the complete treatment of the Coriolis acceleration when using 1D spherical structure models (we refer the reader to their Figures 1 \& 3). In addition, they showed that this latter treatment is also in qualitative agreement with a 2D treatment that takes the full Coriolis and centrifugal accelerations into account\footnote{the centrifugal acceleration is taken into account in the 2D oscillation code ACOR \citep{Ouazzanietal2012} while the 2D rotating stellar model is built by deforming a 1D initial spherical model following the iterative method proposed in \cite{Roxburgh2006}.} (we refer the reader to their Figure 6), but with some quantitative differences that should be explained. Indeed, while the period spacings of individual modes is different, the global properties such as the mean value of the period spacing, the number of modes, the extend and the slope of the pattern for each group of fixed azimuthal order are similar. This could be understood by invoking that studied low-frequency gravito-inertial modes are propagating in (and sounding) deep stellar layers, which are less influenced by the centrifugal acceleration than the surface \citep[e.g.][]{Ballotetal2010}. In this framework, it becomes important to study if the TAR could be generalised to take into account the effects of the centrifugal acceleration in the case of slightly deformed moderately rotating stars (this work) before considering the more extreme cases of stars rotating close to their break-up velocity. This could have several key applications such as new seismic diagnosis, the study of the transport of angular momentum by GIWs \citep[e.g.][]{LeeSaio1993,Mathisetal2008,Mathis2009,Leeetal2014}, and the evaluation of tidal dissipation \citep[][]{OgilvieLin2004,OgilvieLin2007,Bravineretal2014} in deformed stars (and planets).

In this work, we focus on the first goal. First, we introduce in Sect.~\ref{sec:formalism} the formalism already introduced in the literature to study oscillation modes in moderately deformed rotating stars \citep[or planets, e.g.][]{SmeyersDenis1971,Saio1981,Lee1993,LeeBaraffe1995}. We consider the simplest case of a uniform rotation to disentangle the effects of the deformation from those of differential rotation. In Sect.~\ref{sec:TARC}, we show how to generalize TAR in this configuration and in Sect.~\ref{sec:LFGIW} we study the dynamics of corresponding low-frequency GIWs. In Sect.~\ref{sec:SD}, we derive their periods that can be used to probe the chemical composition of stars and their internal rotation. In Sect.~\ref{sec:app}, we use the new formalism to determine how the solutions of the Laplace tidal equation are affected by the centrifugal deformation. Finally, we give in Sect.~\ref{sec:CP} the conclusions of this theoretical work and we discuss its perspectives and future applications.

\section{The dynamical equations in moderately deformed stars}
\label{sec:formalism}

We follow the formalism presented by \cite{LeeBaraffe1995} \citep[see also][]{SmeyersDenis1971,Saio1981,Lee1993} to describe the dynamics of stellar oscillation modes in slightly deformed stars. We work in a spheroidal system of coordinates $\left(a,\theta,\varphi\right)$ to take into account the centrifugal deformation of the star. The origin of this coordinate system ($a=0$) is the center of the deformed star, $\theta$ is the colatitude with $\theta=0$ on the rotation axis, and $\varphi$ is the azimuth. The spheroidal coordinates system is related to the spherical coordinates $\left(r,\theta,\varphi\right)$ by a mapping
\begin{equation}
r=a\left[1+\varepsilon\left(a,\theta\right)\right],
\label{isopot}
\end{equation}
where $\varepsilon$ is a function describing the centrifugal perturbation of the hydrostatic balance. We refer the reader to the Appendix \ref{Appendix:deformation} for its derivation. It scales as $\Omega^2$, where $\Omega$ is the angular velocity of the star, which is here assumed to be uniform. We define the unit-vector basis attached to the spheroidal coordinates system:
\begin{eqnarray}
&&{\widetilde{\bf e}}_{a}=\left(1+\varepsilon+a\partial_{a}\varepsilon\right){\bf \widehat e}_{r},\nonumber\\
&&{\widetilde {\bf e}}_{\theta}=\partial_{\theta}\varepsilon\,{\bf \widehat e}_{r} + \left(1+\varepsilon\right)\,{\bf \widehat e}_{\theta},\nonumber\\
&&{\widetilde {\bf e}}_{\varphi}=\left(1+\varepsilon \right)\,{\bf \widehat e}_{\varphi}.
\end{eqnarray}\\

The dynamical equations for waves are derived in these spheroidal coordinates. As in studies of TAR in spherical stars, we focus in this work on adiabatic oscillations.

First, the linearised Navier-Stockes equation is written
\begin{eqnarray}
\lefteqn{-\omega^2\left[\left(1+2\varepsilon\right){\boldsymbol\xi}+a\,\xi_{a}{\boldsymbol\nabla}_{0}\varepsilon+a\left({\boldsymbol\xi}\cdot{\boldsymbol\nabla}_{0}\varepsilon\right){\widetilde{\bf e}}_{a}\right]+{\vec C}
}\nonumber\\
&=&-{\boldsymbol\nabla}_{0}{\widetilde\Phi}-\frac{1}{{\overline\rho}}{\boldsymbol \nabla}_{0}{\widetilde P}+\frac{{\widetilde\rho}}{{\overline\rho}^2}\frac{{\rm d}{\overline P}}{{\rm d}a}{\widetilde{\bf e}}_{a},\end{eqnarray}
where
\begin{equation}
{\boldsymbol\nabla}_{0}X\equiv\partial_{a}X\,{\widetilde{\bf e}}_{a}+\frac{1}{a}\partial_{\theta}X\,{\widetilde {\bf e}}_{\theta}+\frac{1}{a\sin\theta}\partial_{\varphi}X\,{\widetilde {\bf e}}_{\varphi}
\end{equation}
and
\begin{equation}
\left({\boldsymbol\xi}\cdot{\boldsymbol\nabla}_{0}\right)X\equiv\xi_{a}\partial_{a}X+\frac{\xi_{\theta}}{a}\partial_{\theta}X+\frac{\xi_{\varphi}}{a\sin\theta}\partial_{\varphi}X.
\end{equation}
We introduce the Lagrangian displacement ${\boldsymbol\xi}$ of the oscillation, the gravific potential $\Phi$, the pressure $P$, and the density $\rho$. The scalar quantities ($\Phi,P,\rho$) are expended as the sum of their hydrostatic components (${\overline\Phi},{\overline P,\overline\rho}$) and of their wave fluctuations (${\widetilde\Phi},{\widetilde P},{\widetilde\rho}$). Finally, the Coriolis acceleration operator is given by:
\begin{equation}
{\vec C}=C_{a}{\widetilde{\bf e}}_{a}+C_{\theta}{\widetilde{\bf e}}_{\theta}+C_{\varphi}{\widetilde{\bf e}}_{\varphi},
\end{equation}
where
\begin{eqnarray}
C_{a}&=&-i\omega2\Omega\left(1+2\varepsilon+a\partial_{a}\varepsilon\right)\sin\theta\xi_{\varphi},\nonumber\\
C_{\theta}&=&-i\omega2\Omega\left(1+2\varepsilon+\tan\theta\partial_\theta\varepsilon\right)\cos\theta\xi_{\varphi},\nonumber\\
C_{\varphi}&=&i\omega2\Omega\left(1+2\varepsilon+a\partial_{a}\varepsilon\right)\sin\theta\xi_{a}\nonumber\\
&&+i\omega2\Omega\left(1+2\varepsilon+\tan\theta\partial_\theta\varepsilon\right)\cos\theta\xi_{\theta}.
\end{eqnarray}
We adopt the Cowling approximation in which the perturbations of the gravific potential induced by pulsations $\left({\widetilde\Phi}\right)$ are neglected \citep{Cowling1941}.

Next, the linearised continuity equation is obtained
\begin{eqnarray}
\lefteqn{{\widetilde\rho}+\frac{1}{a^2}\partial_{a}\left(a^2{\overline\rho}\,\xi_{a}\right)+\frac{1}{a\sin\theta}\partial_{\theta}\left(\sin\theta{\overline\rho}\,\xi_{\theta}\right)+\frac{1}{a\sin\theta}\partial_{\varphi}\left({\overline\rho}\,\xi_{\varphi}\right)}\nonumber\\
&+&{\overline\rho}\,{\boldsymbol \xi}\cdot{\boldsymbol\nabla}_{0}\left(3\varepsilon+a\partial_{a}\varepsilon\right)=0.
\label{eq:continuity}
\end{eqnarray}

The linearised energy equation in the adiabatic limit is derived
\begin{equation}
\frac{{\widetilde\rho}}{{\overline\rho}}-\frac{1}{\Gamma_1}\frac{{\widetilde P}}{{\overline P}}+\frac{\xi_a}{a}\left(\frac{{\rm d}\ln{\overline\rho}}{{\rm d}\ln a}-\frac{1}{\Gamma_1}\frac{{\rm d}\ln{\overline P}}{{\rm d}\ln a}\right)=0,
\label{eq:Energy}
\end{equation}
where $\Gamma_1=\left(\partial\,{\ln}\,{\overline P}/\partial\,{\ln}\,{\overline\rho}\right)_S$ ($S$ being the macroscopic entropy) is the adiabatic exponent. It allows us to identify the squared Brunt-Va\"is\"al\"a frequency
\begin{equation}
N^2\left(a\right)=-\frac{\overline g}{a}\left(\frac{{\rm d}\ln{\overline\rho}}{{\rm d}\ln a}-\frac{1}{\Gamma_1}\frac{{\rm d}\ln{\overline P}}{{\rm d}\ln a}\right).
\end{equation}

Finally, the wave's displacement and fluctuations (${\widetilde X}\equiv\left\{{\widetilde\rho},{\widetilde P}\right\}$) are expanded on Fourier series both in time and in azimuth
\begin{eqnarray}
{\boldsymbol \xi}\left(a,\theta,\varphi,t\right)&\equiv&\sum_{\omega,m}\left\{{\boldsymbol{\xi}}^{'}\left(a,\theta\right)\exp\left[i\left(m\varphi+\omega t\right)\right]\right\},\\
{\widetilde X}\left(a,\theta,\varphi,t\right)&\equiv&\sum_{\omega,m}\left\{{X}^{'}\left(a,\theta\right)\exp\left[i\left(m\varphi+\omega t\right)\right]\right\},
\end{eqnarray}
where $\omega$ is the frequency in the co-rotating frame and $m$ the azimutal degree.\\

\section{The TAR including the centrifugal acceleration}
\label{sec:TARC}

We consider each component of the momentum equation to identify the hierarchy of the different terms and the corresponding simplifications when using the TAR in deformed stars.\\

First, the spheroidal radial component can be written
\begin{eqnarray}
\lefteqn{-N^2\left(\frac{\omega}{N}\right)^{2}\left[\left(1+2\left(\varepsilon+a\partial_a\varepsilon\right)\right)\xi_{a}+\xi_{\theta}\partial_{\theta}\varepsilon\right]}\nonumber\\
&&-iN^2\left(\frac{\omega}{N}\right)\left(\frac{2\Omega}{N}\right)\left(1+2\varepsilon+a\partial_a\varepsilon\right)\sin\theta\xi_{\varphi}=-\partial_{a}{\widetilde W}-N^2\xi_{a}\nonumber\\
&&-\frac{1}{{\overline\rho}^2}\partial_{a}{\overline\rho}{\widetilde P},
\label{eq:Tarradial}
\end{eqnarray}
where ${\widetilde W}={\widetilde P}/{\overline\rho}$ and we explicit the frequency ratios $\omega/N$ and $2\Omega/N$. When using TAR, one focuses on low-frequency waves for which $\omega\ll N$. Their propagation can be studied within the anelastic approximation in which acoustic waves are filtered out. Equations (\ref{eq:Tarradial}) and (\ref{eq:Energy}) can be simplified accordingly by neglecting the terms $1/{\overline\rho}^{2}\partial_{a}{\overline\rho}{\widetilde P}$ and $1/\Gamma_{1}\, {\widetilde P}/{\overline P}$, respectively \citep[we refer the reader to sec. 2 \& 3 in][for more details]{Mathis2009}. In addition, TAR can be applied only in the case of "strong" stratification for which $2\Omega\ll N$. In this case, the buoyancy force dominates the radial components of the Coriolis acceleration and of the wave's acceleration. Therefore, the radial momentum equation can be simplified to
\begin{equation}
-\partial_{a}W^{'}-N^2\xi_{a}^{'}=0,
\label{eq:TradRad}
\end{equation}
as in the case of spherical stars and for the same reasons.\\

Next, we examine the latitudinal component of the momentum equation, which we writte
\begin{eqnarray}
\lefteqn{-\omega^2\left[\left(1+2\varepsilon\right)+\left(\frac{\xi_a}{\xi_\theta}\right)\partial_\theta\varepsilon\right]\xi_{\theta}}\nonumber\\
&&-i\omega2\Omega\left(1+2\varepsilon+\tan\theta\,\partial_\theta\varepsilon\right)\cos\theta\xi_{\varphi}=-\frac{1}{a}\partial_{\theta}{\widetilde W}.
\end{eqnarray}
In the case of low-frequency GIWs, the wave displacement is mostly horizontal, i.e. $\xi_a\ll \left\{\xi_{\theta},\xi_{\varphi}\right\}$, because of the strong stable stratification. This allows us to neglect the term $\left(\xi_a/\xi_\theta\right)\partial_\theta\varepsilon$ that couples horizontal and vertical directions. The previous equation simplifies onto
\begin{equation}
-\omega^2{\mathcal A}\left(a,\theta\right)\xi_{\theta}^{'}-i\omega 2\Omega{\mathcal B}\left(a,\theta\right)\cos\theta\xi_{\varphi}^{'}=-\frac{1}{a}\partial_{\theta}W^{'},
\label{TradLat}
\end{equation}
where
\begin{eqnarray}
&&{\mathcal A}=1+2\varepsilon,\\
&&{\mathcal B}={\mathcal A}+\tan\theta\,\partial_{\theta}\varepsilon.
\end{eqnarray}
In the case of a uniform rotation $\varepsilon\left(a,\theta\right)=\varepsilon_{0}\left(a\right)+\varepsilon_{2}\left(a\right)P_{2}\left(\cos\theta\right)$ (see Eq. \ref{epsilon}). Therefore, $\partial_{\theta}\varepsilon\propto\cos\theta\sin\theta$ and the term $\tan\theta\partial_\theta\varepsilon$ is regular.

For the same reasons, the azimuthal component of the momentum equation
\begin{eqnarray}
\lefteqn{-\omega^2\left(1+2\varepsilon\right)\xi_{\varphi}+i\omega2\Omega\left(1+2\varepsilon+a\partial_{a}\varepsilon\right)\sin\theta\xi_{a}}\nonumber\\
&&+i\omega2\Omega\left(1+2\varepsilon+\tan\theta\,\partial_\theta\varepsilon\right)\cos\theta\xi_{\theta}=-\frac{1}{a\sin\theta}\partial_{\varphi}{\widetilde W}
\end{eqnarray}
reduces to
\begin{equation}
-\omega^2{\mathcal A}\left(a,\theta\right)\xi_{\varphi}^{'}+i\omega 2\Omega{\mathcal B}\left(a,\theta\right)\cos\theta\xi_{\theta}^{'}=-\frac{imW^{'}}{a\sin\theta}.
\label{TradAzi}
\end{equation}
As in the spherical case, we thus obtain decoupled equations for the vertical and horizontal components of the displacement. We can thus solve the system formed by Eqs. (\ref{TradLat}) \& (\ref{TradAzi}) and express $\xi_{\theta}^{'}$ and $\xi_{\varphi}^{'}$ as a function of the normalized pressure $W'$: 
\begin{equation}
\xi_{\theta}^{'}=\frac{1}{a}\frac{1}{\omega^2}\frac{1}{{\mathcal D}\left(a,\theta\right)}\left[\partial_{\theta}W^{'}+m\nu\frac{\cos\theta}{\sin\theta}{\mathcal C}\left(a,\theta\right)W^{'}\right],
\end{equation}
\begin{equation}
\xi_{\varphi}^{'}=i\frac{1}{a}\frac{1}{\omega^2}\frac{1}{{\mathcal D}\left(a,\theta\right)}\left[\nu\,{\mathcal C}\left(a,\theta\right)\cos\theta\,\partial_{\theta}W^{'}+\frac{m}{\sin\theta}W^{'}\right],
\end{equation}
with
\begin{eqnarray}
&&{\mathcal C}=\frac{\mathcal B}{\mathcal A}=1+\frac{\tan\theta\,\partial_{\theta}\varepsilon}{1+2\varepsilon},\\
&&{\mathcal D}={\mathcal A}\left(1-\nu^2\cos^2\theta\,{\mathcal C}^{2}\right)\\
&&=\left(1+2\varepsilon\right)\left[1-\nu^2\cos^2\theta\,\left(1+\frac{\tan\theta\,\partial_{\theta}\varepsilon}{1+2\varepsilon}\right)^2\right],\\
&&\nu=\frac{2\Omega}{\omega}.
\end{eqnarray}
We can identify that the structure of the equations in the spheroidal case is very similar to those in the usual spherical case \citep[e.g.][]{LeeSaio1997,Townsend2003}. The most important difference is that the coefficients ${\mathcal A}$, ${\mathcal B}$, ${\mathcal C}$ and ${\mathcal D}$ are function of $a$ and $\theta$ through $\varepsilon\left(a,\theta\right)$ while they reduce to functions that only depend on $\theta$ in the spherical case with ${\mathcal A}={\mathcal B}={\mathcal C}=1$ and ${\mathcal D}=1-\nu\cos^2\theta$. This situation is similar to the case where differential rotation is taken into account \citep[][]{Mathis2009,VanReethetal2018} and we will see in the next section that it will be possible to solve the problem by adopting the same method. We identify as in the uniformly rotating spherical case the so-called spin parameter $\nu=2\Omega/\omega$. The regime $\nu>1$ ($\nu<1$) corresponds to sub- (super-) inertial waves in which $\omega<2\Omega$ ($\omega>2\Omega$).

Following \cite{LeeSaio1997} and \cite{Mathis2009}, we introduce the reduced latitudinal coordinate $x=\cos\theta$; Eqs. (\ref{TradLat}) \& (\ref{TradAzi}) transform onto:
\begin{eqnarray}
\lefteqn{\xi_{\theta}^{'}\left(a,x\right)={\mathcal L}_{\nu m}^{\theta}\left[W^{'}\left(a,x\right)\right]}\nonumber\\
&=&\frac{1}{a}\frac{1}{\omega^2}\frac{1}{\mathcal D\left(a,x\right)}\frac{1}{\sqrt{1-x^2}}\left[-\left(1-x^2\right)\partial_{x}+m\nu x\,{\mathcal C}\left(a,x\right)\right]W^{'},\nonumber\\
\label{eq:tradtheta}
\end{eqnarray}
\begin{eqnarray}
\lefteqn{\xi_{\varphi}^{'}\left(a,x\right)={\mathcal L}_{\nu m}^{\varphi}\left[W^{'}\left(a,x\right)\right]}\nonumber\\
&=&i\frac{1}{a}\frac{1}{\omega^2}\frac{1}{\mathcal D\left(a,x\right)}\frac{1}{\sqrt{1-x^2}}\left[-\nu x\,{\mathcal C}\left(a,x\right)\left(1-x^2\right)\partial_x+m\right]W^{'}.\nonumber\\
\label{eq:tradphi}
\end{eqnarray}

\section{The dynamics of low-frequency gravito-inertial waves}
\label{sec:LFGIW}
As in \cite{Mathis2009} and \cite{VanReethetal2018}, we focus from now on low-frequency GIWs. Our goal is to derive the so-called Poincar\'e equation for the normalized pressure $\left(W^{'}\right)$ that will allow us to compute their frequencies and periods and to built the corresponding seismic diagnosis. Using again the anelastic approximation, the continuity equation becomes ${\boldsymbol\nabla}\cdot\left({\overline\rho}{\vec u}\right)=0$, where we recall the relation ${\vec u}=\partial_t\,{\boldsymbol\xi}=i\omega\,{\boldsymbol\xi}$ between the velocity $\left({\vec u}\right)$ and the Lagrangian displacement $\left({\boldsymbol\xi}\right)$. The continuity equation (Eq. \ref{eq:continuity}) simplifies onto
\begin{eqnarray}
\lefteqn{\frac{1}{a^2}\partial_{a}\left(a^2{\overline\rho}\,\xi_{a}^{'}\right)+\frac{1}{a\sin\theta}\partial_{\theta}\left(\sin\theta{\overline\rho}\,\xi_{\theta}^{'}\right)+\frac{1}{a\sin\theta}\partial_{\varphi}\left({\overline\rho}\,\xi_{\varphi}^{'}\right)}\nonumber\\
&+&{\overline\rho}\,\left[\xi_{a}^{'}\partial_a{\mathcal E}+\frac{\xi_\theta^{'}}{a}\partial_{\theta}{\mathcal E}\right]=0,
\label{eq:anelastic}
\end{eqnarray}
where
\begin{equation}
{\mathcal E}=3\varepsilon+a\partial_{a}\varepsilon.
\end{equation}
Low-frequency GIWs are rapidly oscillating with short wavelengths along the (vertical) ${\widetilde{\bf e}}_{a}$ direction that are very small compared to the characteristic lengths of variation of the background quantities. This allows us following \cite{Mathis2009} to use the Jeffreys-Wentzel-Kramers-Brillouin (hereafter JWKB) approximation \citep{FromanFroman2005} along the vertical and to expand the pressure fluctuation and the components of the displacement as
\begin{eqnarray}
W^{'}&=&\sum_{k}\left\{w_{\nu k m}\left(a,\theta\right)\frac{A_{\nu k m}}{k_{V;\nu k m}^{1/2}}\exp\left[i\int^{a}k_{V;\nu k m}{\rm d}a\right]\right\}\\
\xi_{j}^{'}&=&\sum_{k}\left\{{\widehat\xi}_{j;\nu k m}\left(a,\theta\right)\frac{A_{\nu k m}}{k_{V;\nu k m}^{1/2}}\exp\left[i\int^{a}k_{V;\nu k m}{\rm d}a\right]\right\},
\end{eqnarray} 
where $j\equiv\left\{r,\theta,\varphi\right\}$, $k$ is the index of a latitudinal eigenmode (see Eq. \ref{tidal}), and $A_{\nu k m}$ is the amplitude of the wave. Using Eqs. (\ref{eq:TradRad}), (\ref{eq:tradtheta}), and (\ref{eq:tradphi}), we can obtain the following polarisation relations:
\begin{eqnarray}
{\widehat\xi}_{r;\nu k m}\left(a,\theta\right)&=&-i\frac{k_{V;\nu k m}}{N^2}w_{\nu k m}\left(a,\theta\right),\nonumber\\
{\widehat\xi}_{\theta;\nu k m}\left(a,\theta\right)&=&{\mathcal L}_{\nu m}^{\theta}\left[w_{\nu k m}\left(a,\theta\right)\right],\nonumber\\
{\widehat\xi}_{\varphi;\nu k m}\left(a,\theta\right)&=&{\mathcal L}_{\nu m}^{\varphi}\left[w_{\nu k m}\left(a,\theta\right)\right].
\label{polarisation}
\end{eqnarray}

Using the vertical momentum equation Eq. (\ref{eq:TradRad}), Eq. (\ref{eq:anelastic}) becomes 
\begin{equation}
\frac{k_{V;\nu k m}^2}{N^2}w_{\nu k m}+\frac{1}{a\sin\theta}\partial_{\theta}\left(\sin\theta{\widehat\xi}_{\theta;\nu k m}\right)+\frac{{\widehat\xi}_{\theta;\nu k m}}{a}\partial_{\theta}{\mathcal E}+\frac{i m {\widehat\xi}_{\varphi;\nu k m}}{a\sin\theta}=0,
\end{equation}
where the JWKB approximation allows us to neglect ${\overline\rho}\,\xi_{a}^{'}\partial_a{\mathcal E}$ in front of the dominant term $\displaystyle{\frac{1}{a^2}\partial_{a}\left(a^2{\overline\rho}\,\xi_{a}^{'}\right)}$.

\begin{table*}[ht!]
\begin{center}
\caption{Centrifugal terms involved in the generalised Laplace tidal equation.}
\begin{tabular}{ c c }
\hline
\hline
${\mathcal A}$ & $1+2\varepsilon$ \\
\\
${\mathcal B}$ & $1+2\varepsilon+\tan\theta\,\partial_{\theta}\varepsilon=\displaystyle{1+2\varepsilon-\frac{1-x^2}{x}\partial_x\varepsilon}$ \\
\\
${\mathcal C}$ &  $\displaystyle{1+\frac{\tan\theta\,\partial_{\theta}\varepsilon}{1+2\varepsilon}}=1-\frac{1}{1+2\varepsilon}\frac{1-x^2}{x}\partial_x\varepsilon$\\
\\
${\mathcal D}$ & $\displaystyle{\left(1+2\varepsilon\right)\left[1-\nu^2\cos^2\theta\,\left(1+\frac{\tan\theta\,\partial_{\theta}\varepsilon}{1+2\varepsilon}\right)^2\right]}=\displaystyle{\left(1+2\varepsilon\right)\left[1-\nu^2 x^2\,\left(1-\frac{1}{1+2\varepsilon}\frac{1-x^2}{x}\partial_x\varepsilon\right)^2\right]}$\\
\\
${\mathcal E}$ & $3\varepsilon+a\partial_{a}\varepsilon$\\
\hline
\end{tabular}
\label{tab:coefficient}
\end{center}
\end{table*}

Using Eqs. (\ref{eq:tradtheta}) and (\ref{eq:tradphi}) allows us to obtain the equation for $w_{\nu k m}$
\begin{eqnarray}
\lefteqn{{\mathcal L}_{\nu m}\left[w_{\nu k m}\right]=\partial_{x}\left[\frac{\left(1-x^2\right)}{\mathcal D}\partial_{x}\,w_{\nu k m}\right]+\frac{\left(1-x^2\right)\partial_{x}{\mathcal E}}{\mathcal D}\partial_{x} w_{\nu k m}}\nonumber\\
&&-\left[\frac{m^2}{\left(1-x^2\right){\mathcal D}}+m\nu\frac{{\rm d}}{{\rm d}x}\left[\frac{x\,{\mathcal 
C}}{\mathcal D}\right]+m\nu\frac{x\,{\mathcal C}}{\mathcal D}\partial_{x}{\mathcal E}\right]w_{\nu k m}\nonumber\\
&&=-\Lambda_{\nu km}\left(a\right)w_{\nu k m},
\label{tidal}
\end{eqnarray}
where we identify the dispersion relation for low-frequency GIWs within TAR
\begin{equation}
k_{V;\nu k m}^2=\frac{N^2\left(a\right)}{\omega_{k m}^{2}}\frac{\Lambda_{\nu k m}\left(a\right)}{a^2}.
\label{Eq:dispers}
\end{equation}
This equation is in fact the Poincar\'e Partial Differential Equation for GIWs. The combined use of the TAR and JWKB approximations, allows us to transform it in a linear second-order Ordinary Differential Equation on $x$ with only a parametric dependence on $a$. This result is very similar to those obtained in the case of differential rotation in \cite{OgilvieLin2004}, \cite{Mathis2009} and \cite{VanReethetal2018}. Therefore, Eq. (\ref{tidal}) can be seen as a generalised Laplace tidal equation (and operator) for generalised Hough functions when taking the centrifugal acceleration into account. In Tab. \ref{tab:coefficient}, we recall the expressions of all the involved coefficients. Their first-order linearisation in $\varepsilon$ is derived in Appendix \ref{sec:fixed}. Since the eigenvalues and eigenfunctions of the generalised Laplace tidal equation vary with the pseudo-radius, we choose to define our latitudinal ordering number $k$ as in \cite{LeeSaio1997} by considering the eigenvalues and eigenfunctions at the center where they are not affected by the centrifugal acceleration since the mapping we choose here (see Eq. \ref{shape} in the Appendix \ref{Appendix:deformation}) is such that $\varepsilon\rightarrow0$ for $a\rightarrow0$. We recall that \cite{LeeSaio1997} ordered the eigenvalues of the non-deformed Hough functions as follows: for the eigenvalues that exist for any value of $\nu$, they attached positive $k$ including zero with $\Lambda_{\nu=0,km}\left(a=0\right)=\left(\vert m \vert+k\right)\left(\vert m\vert+k+1\right)$ that corresponds to the $l\left(l+1\right)$ eigenvalues of the spherical harmonics (with $l=\vert m \vert +k$), which are the horizontal eigenfunctions for non-radial pulsations in the non-rotating case (i.e. $\nu=0$); for the eigenvalues that only exist when $\vert \nu \vert>1$, they used negative $k$ in such a way that $\Lambda_{\nu,-1,m}\left(a=0\right)>\Lambda_{\nu,-2,m}\left(a=0\right)>...$. As we will see in the next section, avoided crossings can appear when looking at the variation of the eigenvalues along the pseudo-radius (see Fig. \ref{fig:spec_r_2}). The corresponding $k$ as defined by \cite{LeeSaio1997} then changes.

As in the uniformly and differentially rotating spherical cases, two classes of waves are identified: those for which ${\mathcal D}>0$ in the whole spheroidal shell and that propagate at all colatitudes and those for which ${\mathcal D}$ vanishes in the spheroidal cavity at a critical colatitude $\theta_c$ such that ${\mathcal D}\left(a,\theta_c\right)=0$, which depends on $a$. For this second class, waves propagate only within an equatorial belt where $\theta>\theta_c\left(a\right)$ \citep[some examples of these waves can be found in][while an approximate value for $\theta_c\left(a\right)$ is derived in Eq. \ref{CritLatCA}]{Ballotetal2010,Pratetal2016,Pratetal2018}. These first and second classes of waves correspond to the super-inertial ($\omega>2\Omega$) and sub-inertial waves ($\omega<2\Omega$), respectively, in the case of the uniformly rotating spherical case. 

This description within the TAR of low-frequency GIWs propagating in deformed bodies can be applied to the study of the transport of angular momentum they induce \citep{Mathis2009}, of tidal dissipation in stably stratified stellar and planetary layers \citep{OgilvieLin2004,OgilvieLin2007,Bravineretal2014,Fulleretal2016}, and of the seismology of rapidly rotating stars \citep[e.g.][]{VanReethetal2018}.

In this framework, it is finally interesting to derive the asymptotic frequencies of low-frequency GIWs and the corresponding periods as in \cite{VanReethetal2018}. Indeed, these latters allow asteroseismologist to probe the internal rotation of regions where these waves propagate, for instance the radiative layers close to the convective core of intermediate-mass stars \citep{VanReethetal2016,Ouazzanietal2017,VanReethetal2018}.

\section{Asymptotic seismic diagnosis}
\label{sec:SD}

\begin{figure}[t!]
\begin{center}
\includegraphics[scale=0.475]{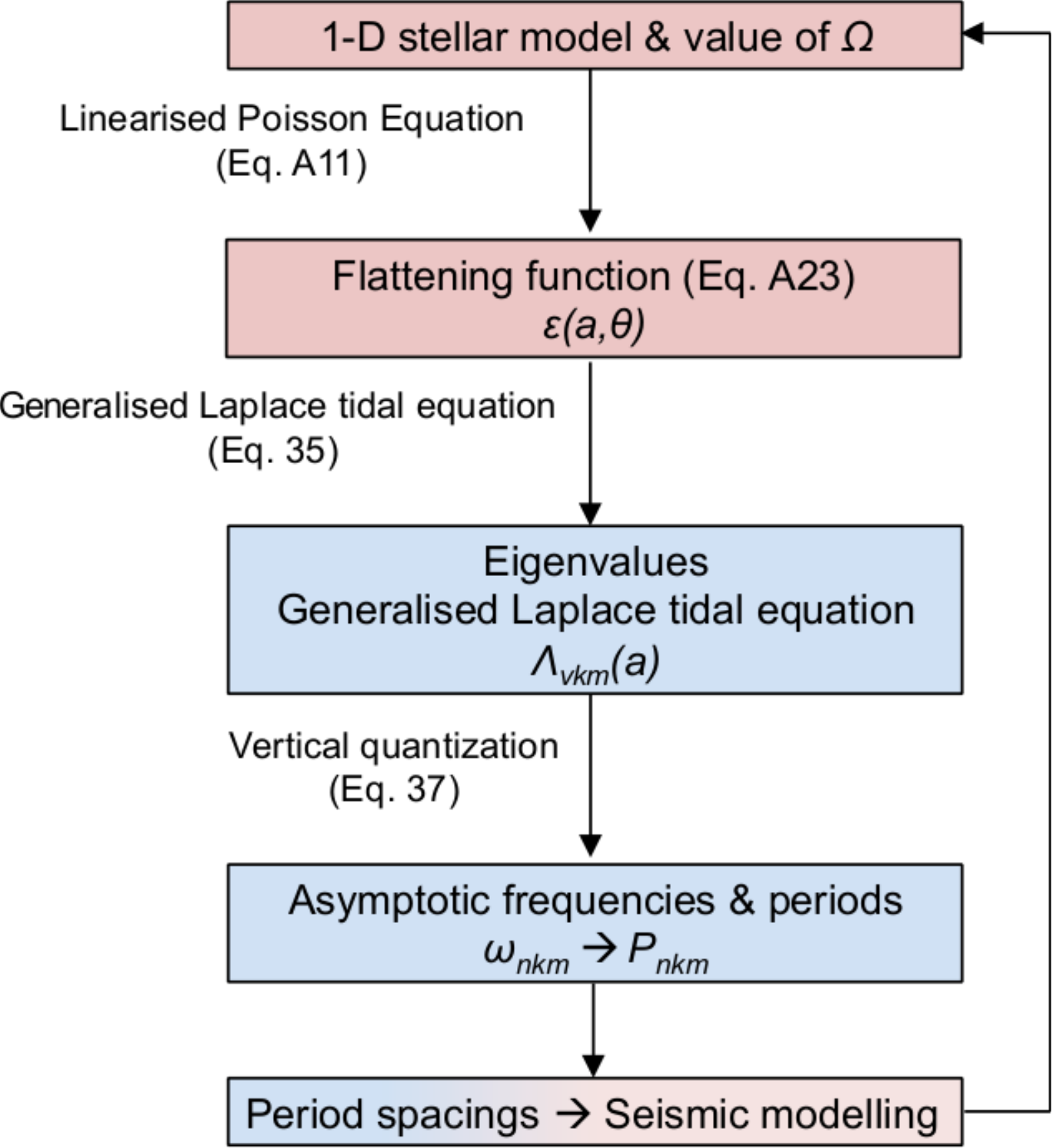}
\end{center}
\caption{Methodology for seismic modeling of rotating deformed stars using the TAR. Red and blue boxes are related to the structure of stars and low-frequency GIWs, respectively.} 
\label{Fig1}       
\end{figure}

Following the same method that is used in the case of spherical uniformly and differentially rotating stars, we can derive the eigenfrequencies of low-frequency GIWs by doing a vertical (i.e., here along $a$) quantization:
\begin{equation}
    \label{eq:quant}
\int_{a_{\rm t1}}^{a_{\rm t2}}k_{V;\nu n k m}{\rm d}a=\frac{1}{\omega_{n k m}}\int_{a_{\rm t1}}^{a_{\rm t2}}\frac{\Lambda^{1/2}_{\nu k m}\left(a\right)N\left(a\right)}{a}{\rm d}a=\left(n+1/2\right)\pi,
\end{equation} 
where we have introduced $n$ the vertical order while $a_{\rm t1}$ and $a_{\rm t2}$ are the inner and outer turning point for which the Brunt-V\"ais\"al\"a frequency ($N$) vanishes \citep{Berthomieuetal1978,Tassoul1980,Bouabidetal2013}.
Using the previously derived dispersion relation (Eq. \ref{Eq:dispers}), we get the asymptotic expression for the frequencies of low-frequency GIWs
\begin{equation}
\omega_{nkm}=\frac{\displaystyle{\int_{a_{\rm t1}}^{a_{\rm t2}}\frac{\Lambda^{1/2}_{\nu k m}\left(a\right)N\left(a\right)}{a}{\rm d}a}}{\left(n+1/2\right)\pi}
\label{eigenfrequencies}
\end{equation}
and the corresponding period
\begin{equation}
P_{nkm}=\frac{2\pi^2\left(n+1/2\right)}{\displaystyle{\int_{a_{\rm t1}}^{a_{\rm t2}}\frac{\Lambda^{1/2}_{\nu k m}\left(a\right)N\left(a\right)}{a}{\rm d}a}}.
\label{eigenperiods}
\end{equation}
As in the spherical case, we can thus compute the period spacing $\Delta P=P_{n k m}-P_{n-1 k m}$ that will allow to probe the internal rotation of stars but with taking into account the flattening of stars by the centrifugal acceleration. The slight differences are the use of spheroidal coordinates and the variation with $a$ of the eigenvalues $\Lambda_{\nu k m}$ of the generalised Laplace tidal operator. 

The interest of the equations derived for the star's deformation (Eq. \ref{shape}), the generalised Hough functions (Eq. \ref{tidal}), and the eigen-frequencies and periods (Eqs. \ref{eigenfrequencies} and \ref{eigenperiods}) is that they can be easily integrated for a large number of stars. This is a great asset when one has to compute complete grids of stellar models to perform detailed seismic modellings \citep[e.g.][and Fig. \ref{Fig1}]{Pedersenetal2018}.

\begin{figure}
    \centering
    \resizebox{\hsize}{!}{\includegraphics{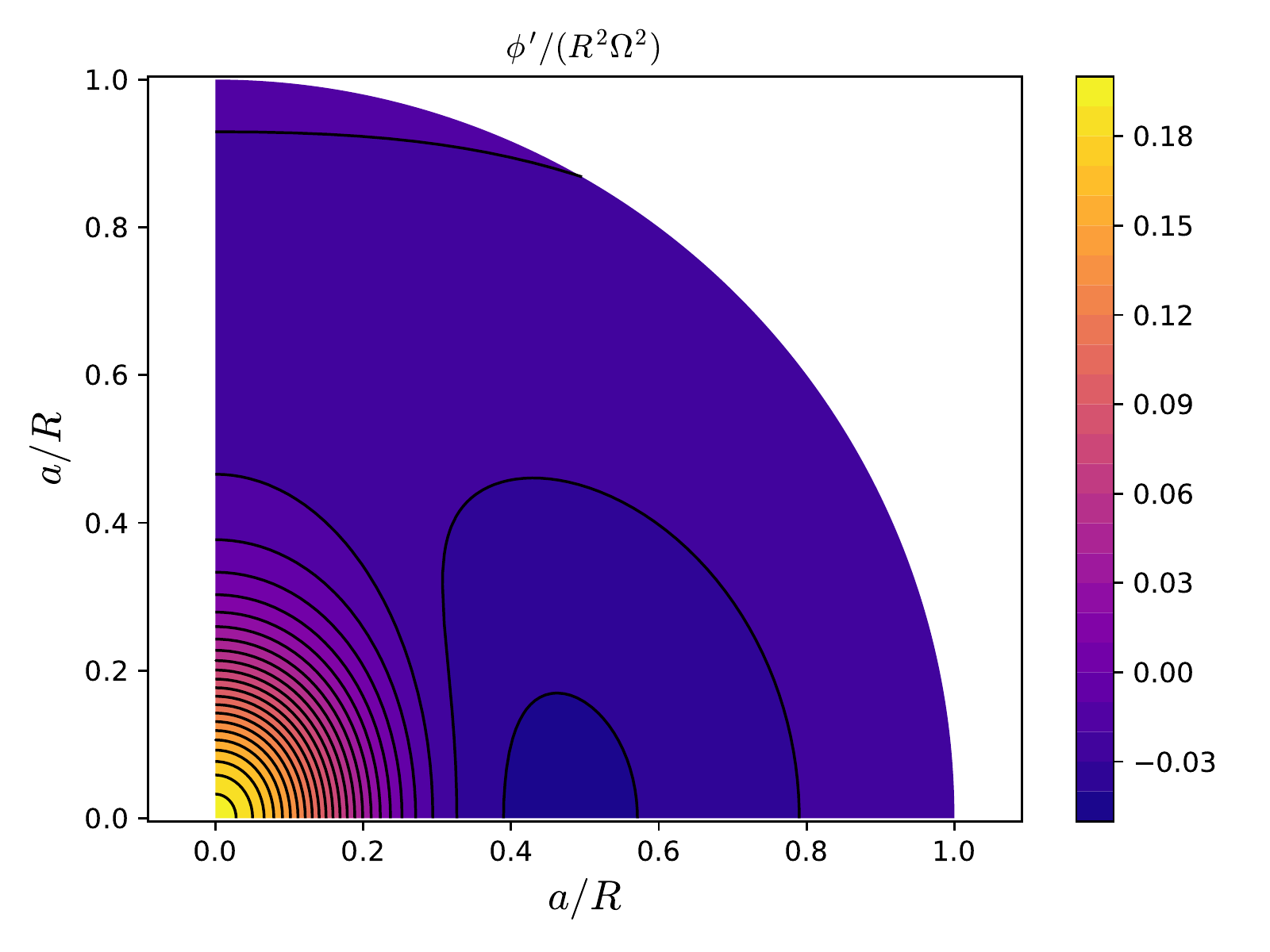}}
    \caption{Normalised pertubation of the gravitational potential $\phi'(a,\theta)/\left(R^2\Omega^2\right)$.}
    \label{fig:phip}
\end{figure}

\begin{figure}
    \centering
    \resizebox{\hsize}{!}{\includegraphics{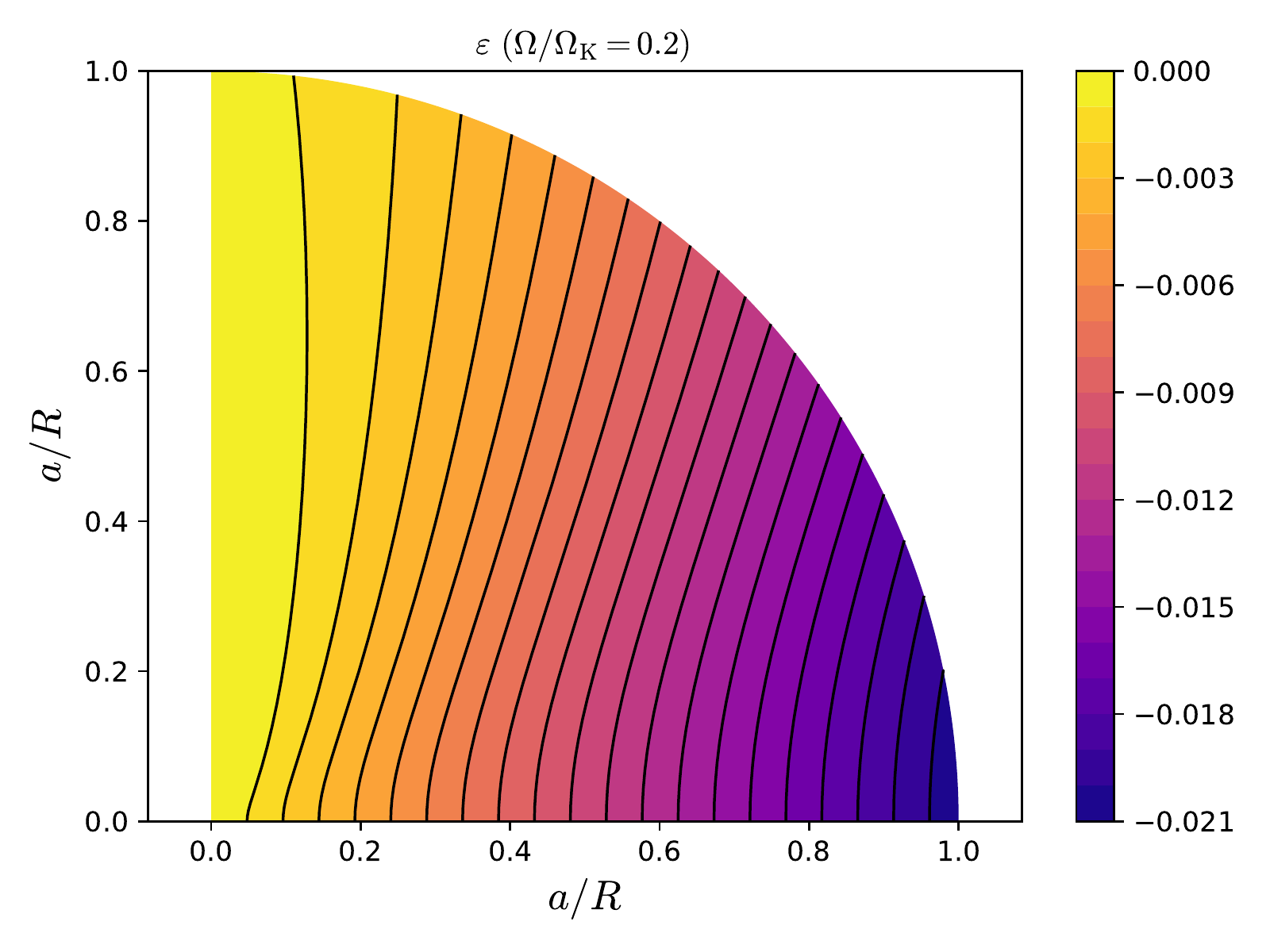}}
    \caption{Deformation function $\varepsilon(a,\theta)$ for $\Omega/\Omega_{\rm K}=0.2$.}
    \label{fig:eps02}
\end{figure}

\begin{figure}
    \centering
    \resizebox{\hsize}{!}{\includegraphics{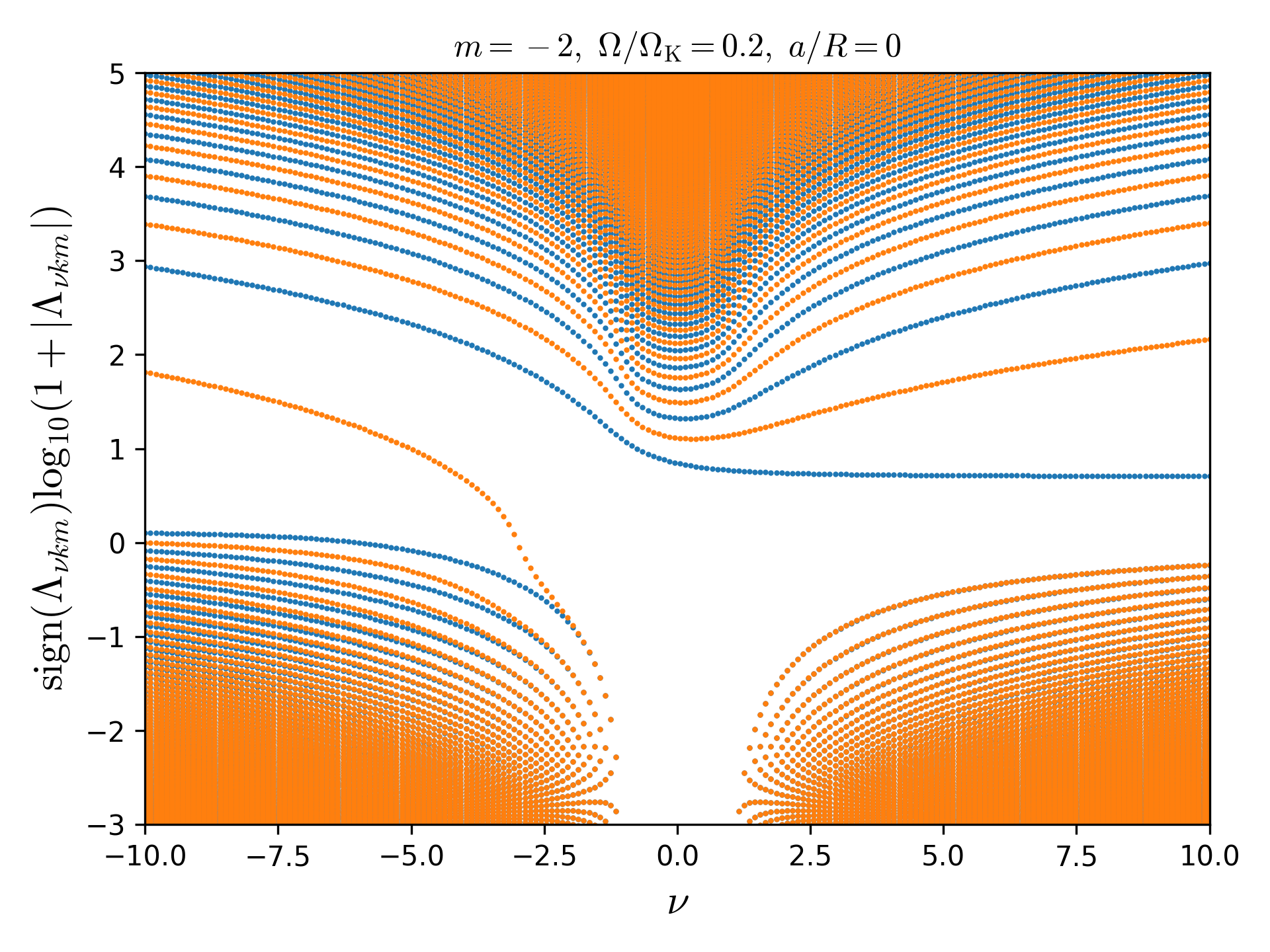}}
    \caption{Spectrum of the generalised Laplace tidal equation as a function of the spin factor $\nu$ at $a=0$ for $\Omega=0.2\Omega_{\rm K}$ and $m=-2$.
Blue (respectively orange) dots correspond to even (respectively odd) eigenfunctions.}
    \label{fig:spec_nu}
\end{figure}

\begin{figure*}[!t]
    \centering
    \resizebox{0.49\hsize}{!}{\includegraphics{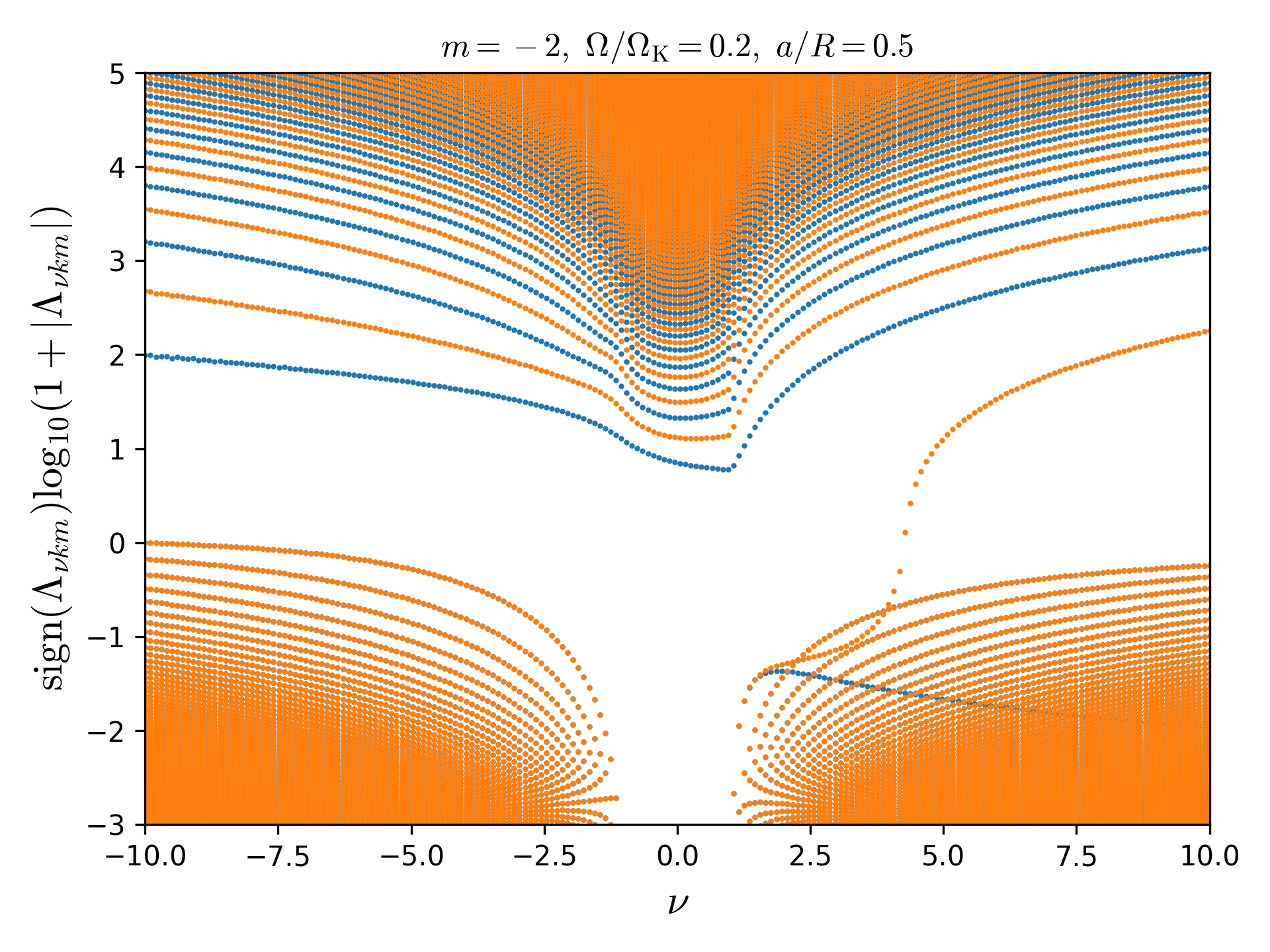}}
    \hfill
    \resizebox{0.49\hsize}{!}{\includegraphics{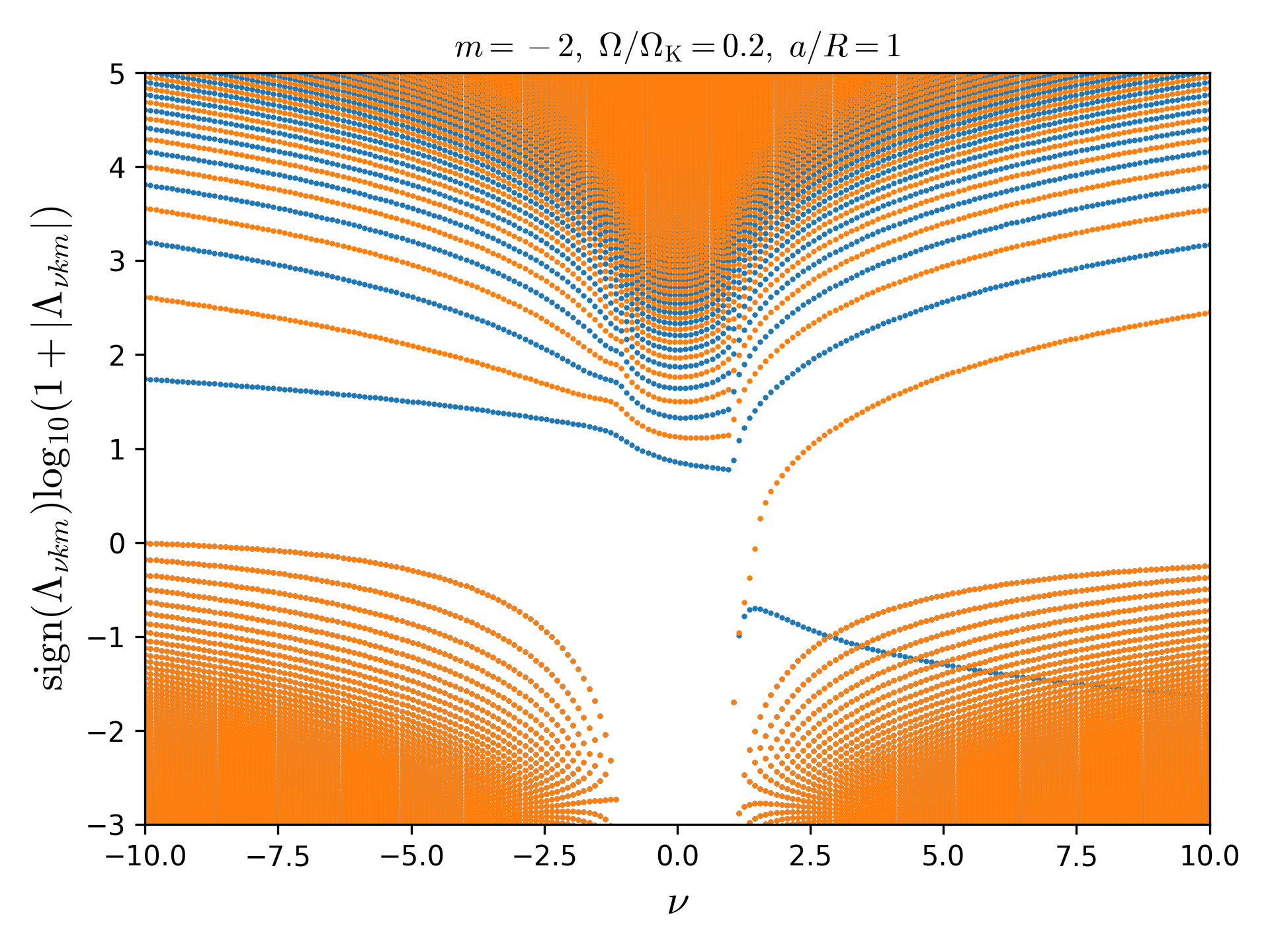}}
    \caption{Same as Fig.~\ref{fig:spec_nu}, but at different pseudo-radii: $a=0.5R$ (left) and $a=R$ (right).}
    \label{fig:spec_nu_02}
\end{figure*}

\begin{figure*}
    \centering
    \resizebox{0.49\hsize}{!}{\includegraphics{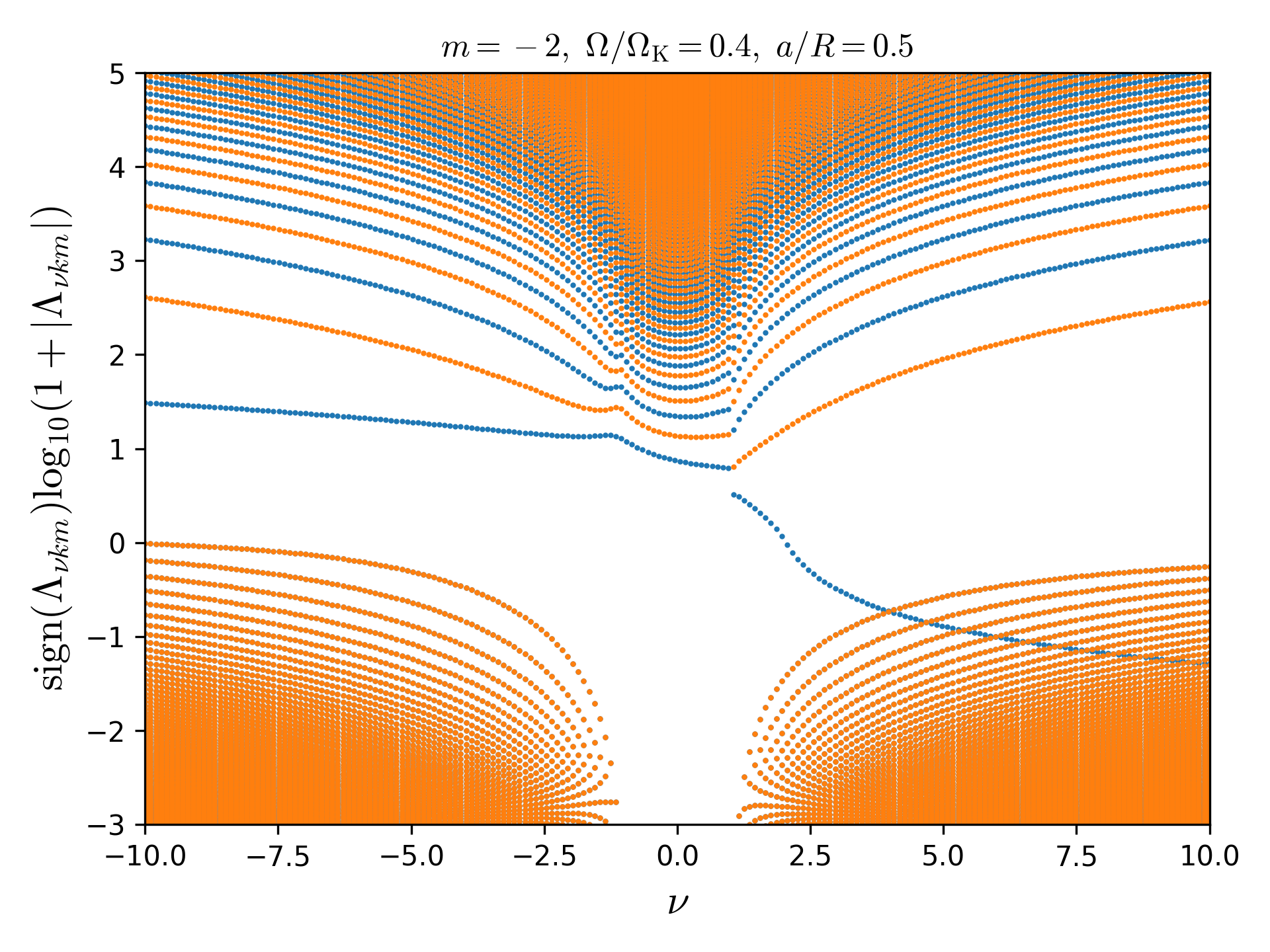}}
    \hfill
    \resizebox{0.49\hsize}{!}{\includegraphics{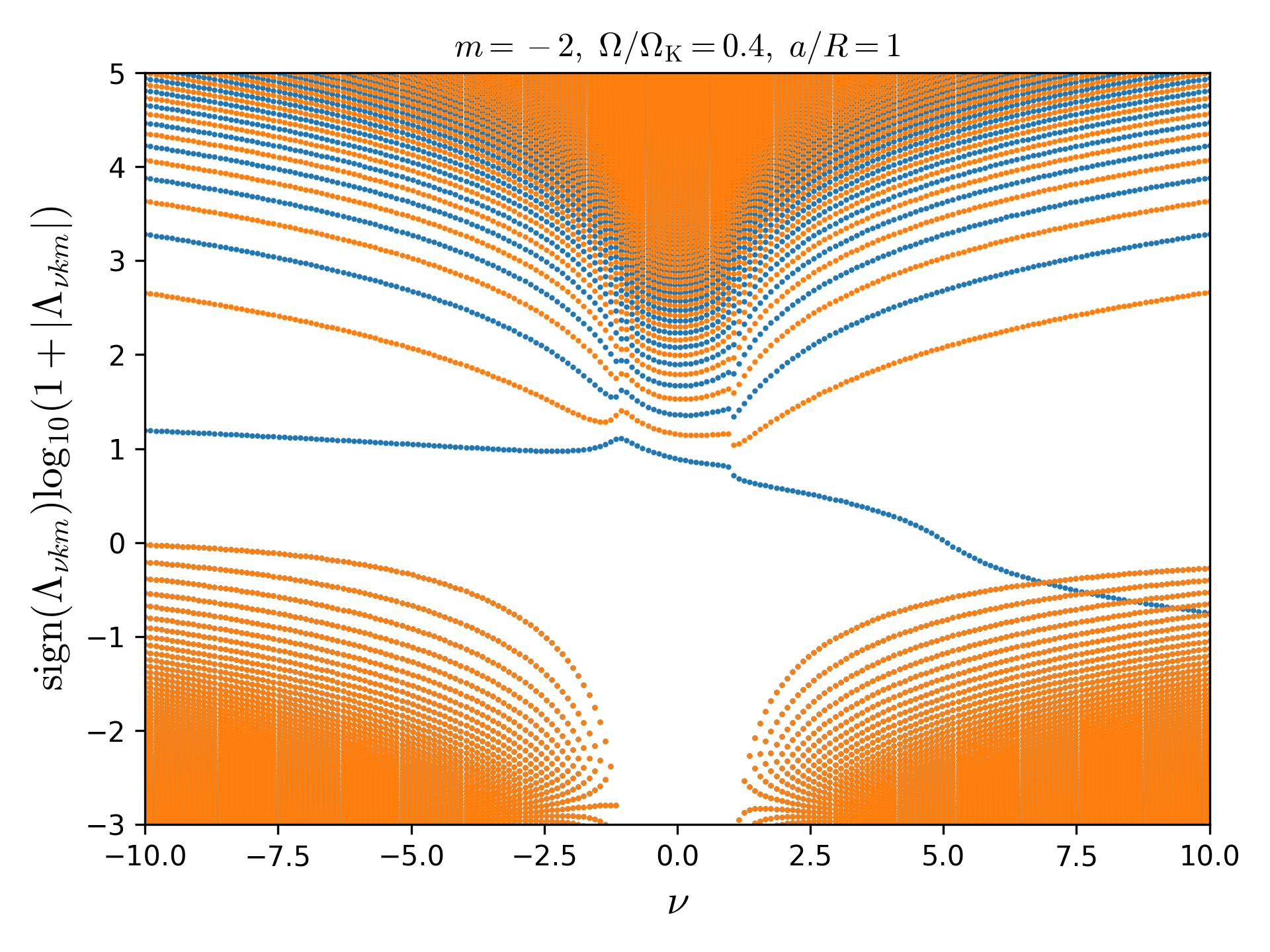}}
    \caption{Same as Fig.~\ref{fig:spec_nu}, but for $\Omega=0.4\Omega_{\rm K}$ and at different pseudo-radii:  $a=0.5R$ (left) and $a=R$ (right).}
    \label{fig:spec_nu_04}
\end{figure*}

\begin{figure*}
    \centering
    \resizebox{0.49\hsize}{!}{\includegraphics{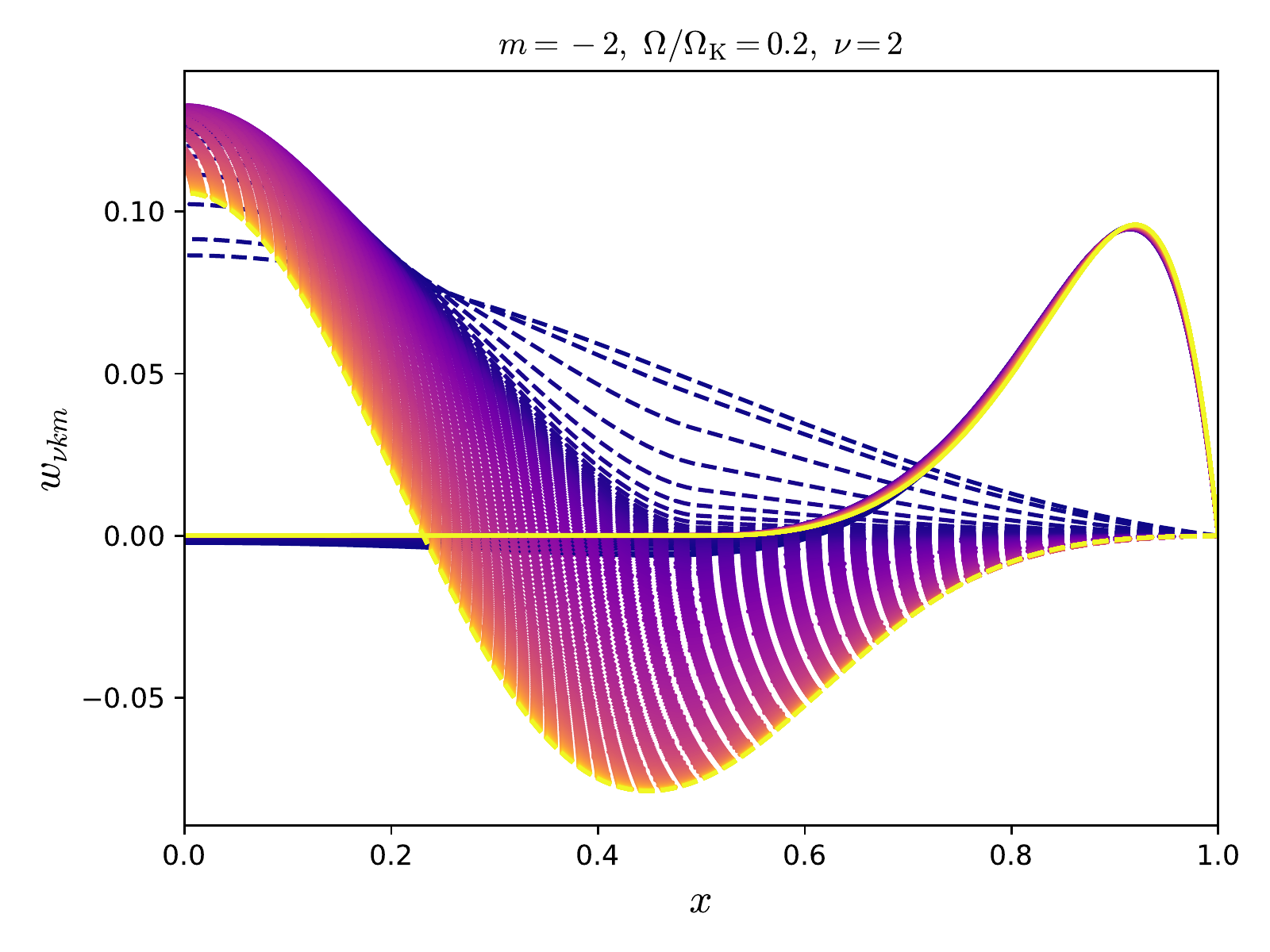}}
    \hfill
    \resizebox{0.49\hsize}{!}{\includegraphics{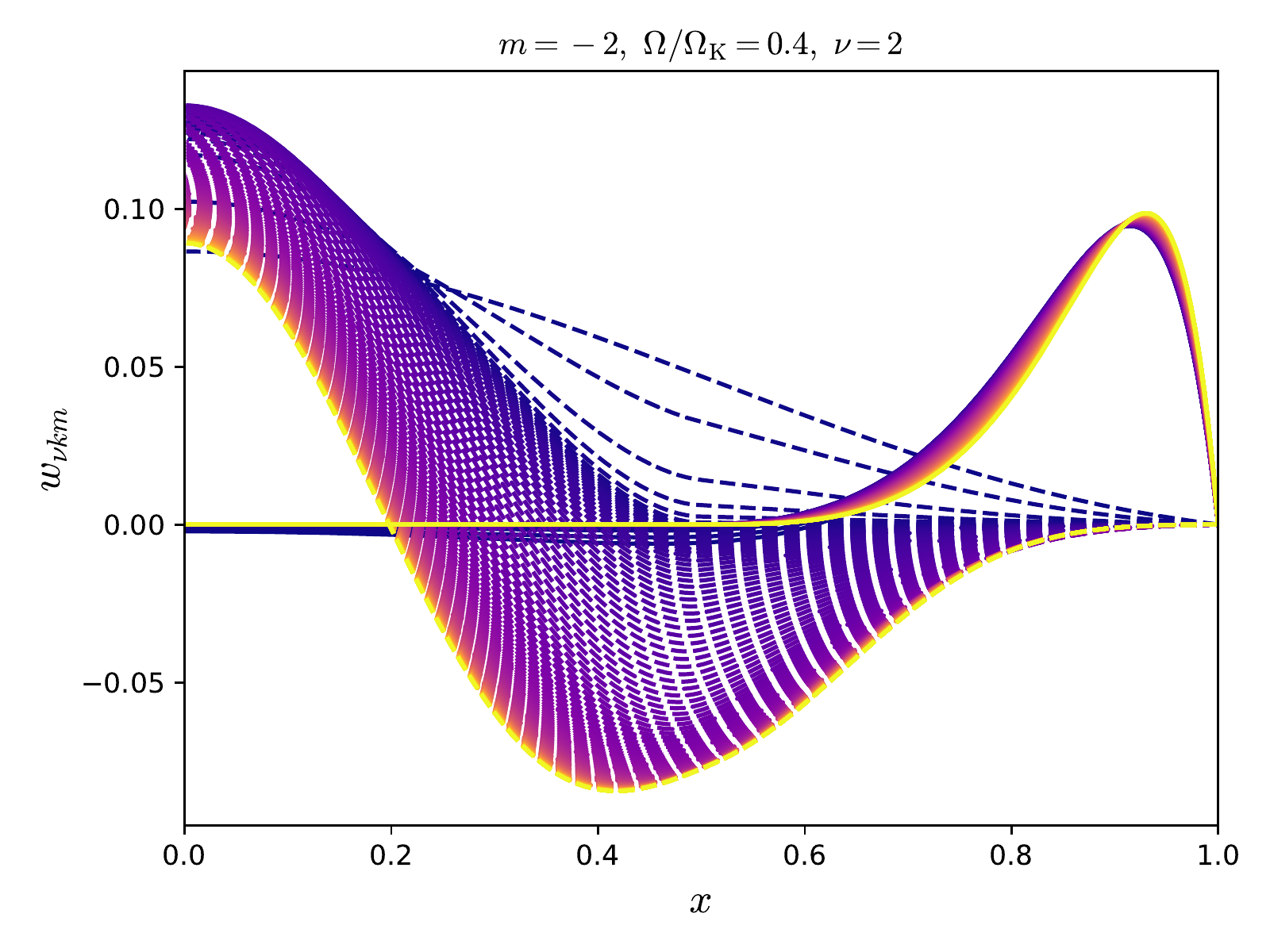}}
    \caption{Solutions of the generalised Laplace tidal equation at different pseudo-radii from $a=0$ (dark blue) to $a=R$ (yellow) for $m=-2$, $\nu=2$ and $\Omega/\Omega_{\rm K}=0.2$ (left) and $\Omega/\Omega_{\rm K}=0.4$ (right).
Dashed lines correspond to gravity-like solutions with $k=0$ (and an avoided crossing with the $k=2$ mode), whereas solid lines correspond to Rossby-like solutions with $k=-4$ (and an avoided crossing with the $k=-2$ mode).}
    \label{fig:func_nu2}
\end{figure*}

\begin{figure*}
    \centering
    \resizebox{0.49\hsize}{!}{\includegraphics{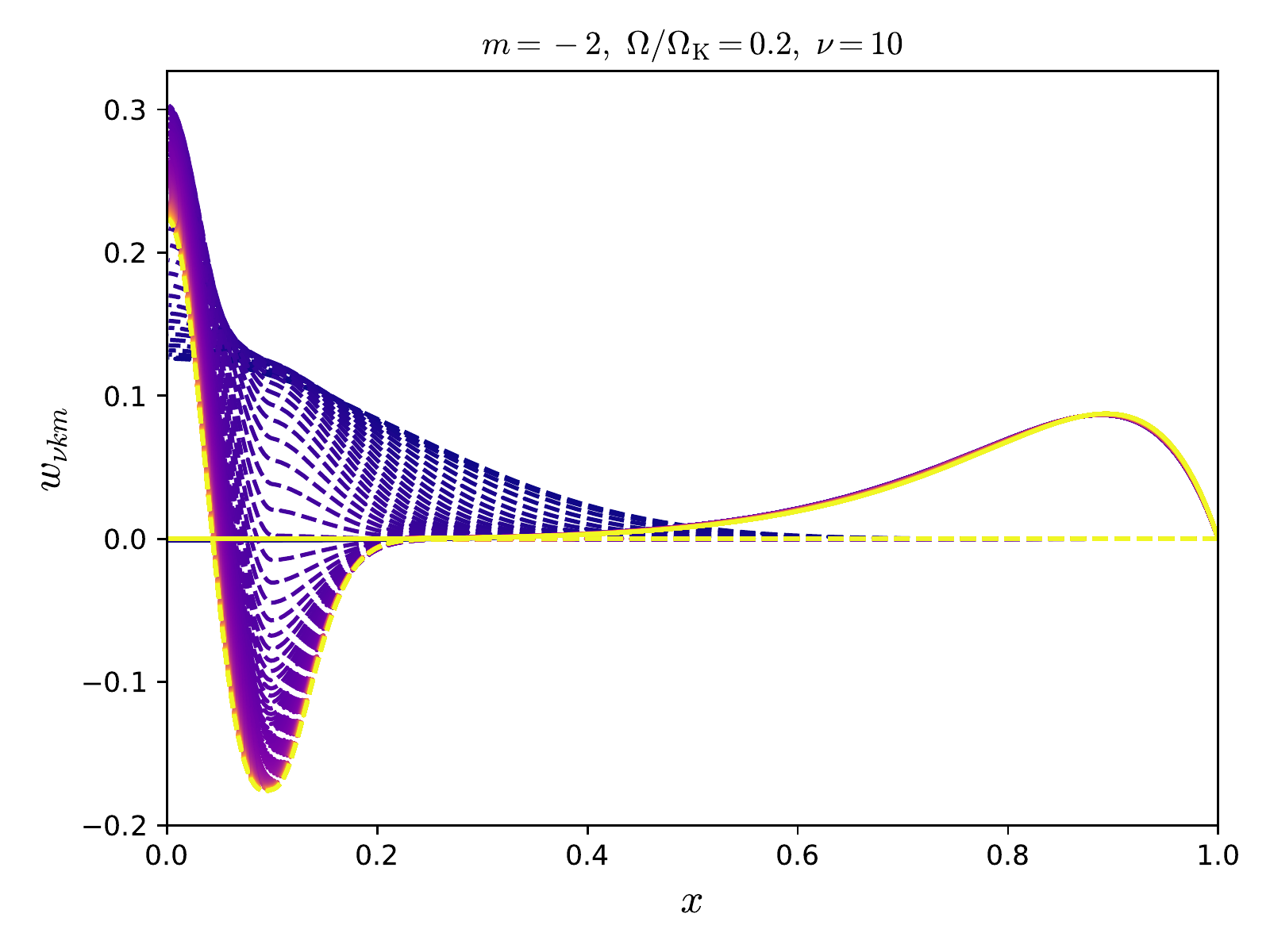}}
    \hfill
    \resizebox{0.49\hsize}{!}{\includegraphics{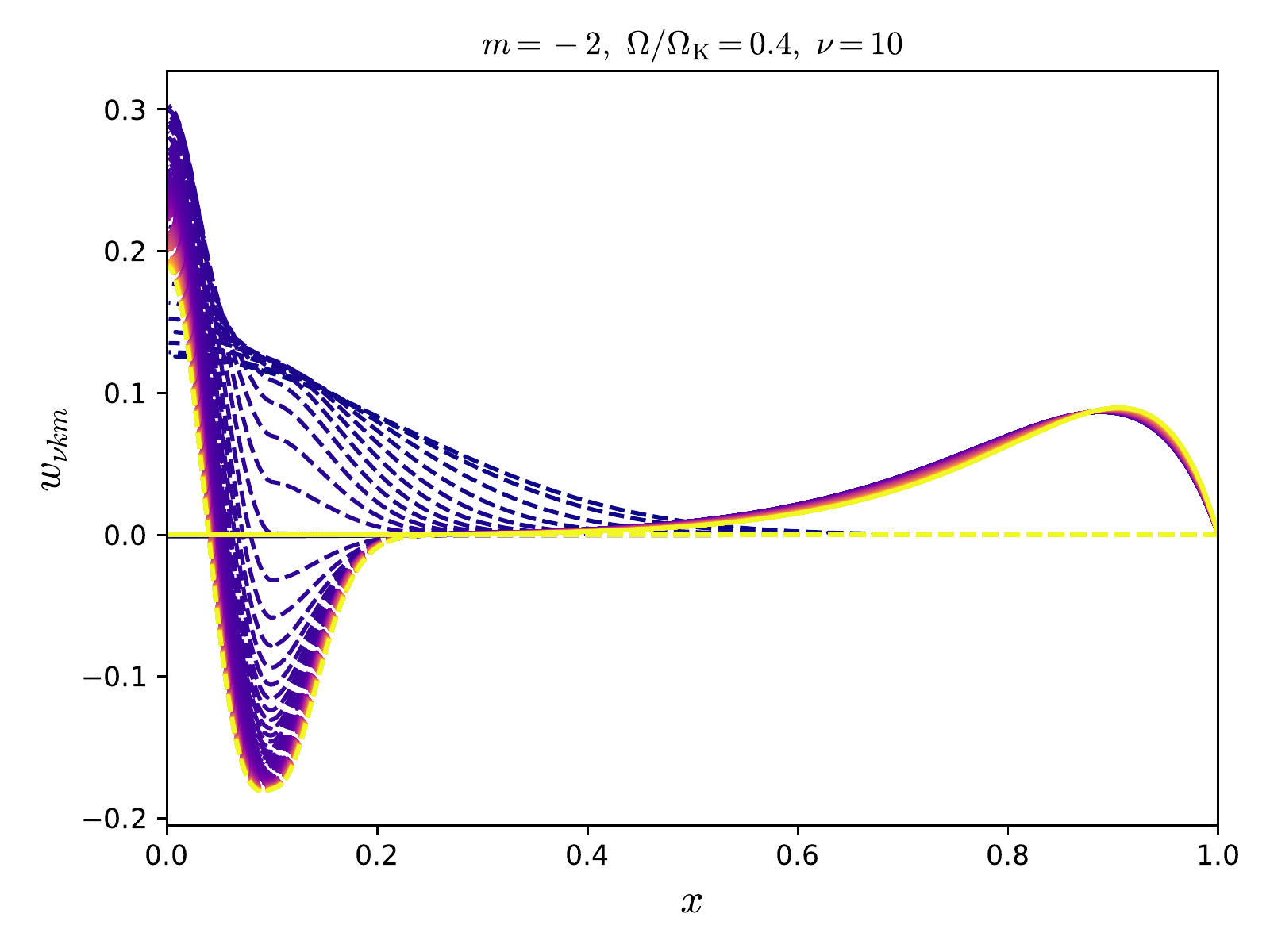}}
    \caption{Same as Fig.~\ref{fig:func_nu2}, but for $\nu=10$.}
    \label{fig:func_nu10}
\end{figure*}

\section{Application}
\label{sec:app}

In this section, we solve numerically the linearized generalised Laplace tidal equation at fixed frequency, as presented in Appendix~\ref{sec:fixed}.
We consider here a ZAMS $1.5\,{\rm M_\odot}$ stellar model with a solar metallicity computed with the MESA 1D stellar evolution code \citep{Paxtonetal2018}.
First, we compute $\varepsilon$ from the perturbation of the gravitational potential $\phi'$, following the procedure described in Appendix~\ref{Appendix:deformation}.
Equation~\eqref{poisson-pert} implies that $\phi'$ is proportional to $\Omega^2$ and can thus be normalized by $R^2\Omega^2$, as represented in Fig.~\ref{fig:phip}.
The deformation function $\varepsilon(a,\theta)$ is illustrated in Fig.~\ref{fig:eps02} for $\Omega/\Omega_{\rm K}=0.2$.
For this rotation rate, its maximum absolute value is of the order of 2\%. In addition, we observe that $\varepsilon<0$. This can be easily understood since by definition (see Eq. \ref{isopot}) $r=a\left(1+\varepsilon\right)$, where $r$ is the usual spherical radius and $a$ the pseudo-radius with $a>r$ because of the action of the centrifugal acceleration. Since $\varepsilon$ scales with the square of the rotation rate, this value is slightly lower than 10\% for $\Omega/\Omega_{\rm K}=0.4$.
A priori, the perturbative approach is thus valid for these rotation rates.

We then solve the generalised Laplace tidal equation for different pseudo-radii, spin factors, and rotation rates using an implementation based on Chebyshev polynomials similar to the one by \cite{Wangetal2016}.
At the center ($a=0$), this is equivalent to solving the unperturbed classical Laplace tidal equation.
The corresponding spectrum as a function of the spin factor is shown for $m=-2$ in Fig.~\ref{fig:spec_nu}.
As expected, it is consistent with Fig.~1 of \citet{LeeSaio1997}.
It features Rossby-like solutions, with mostly negative eigenvalues and inexistent for $|\nu|<1$, and gravity-like solutions, with positive eigenvalues.
A notable property of the Rossby-like part of the spectrum for $\nu>1$ is that every odd solution can be associated with an even solution that has a very close eigenvalue.

Spectra at different pseudo-radii (illustrated in Fig.~\ref{fig:spec_nu_02}) show significant effects of the centrifugal deformation.
First, Rossby-like solutions for $\nu<-1$ are now also grouped by two (one odd solution and one even solution) with very close eigenvalues.
Second, two previously grouped Rossby-like solutions for $\nu>1$ now have very different behaviors, and probably cause avoided crossings with other Rossby-like solutions.
Finally, the behavior of gravity-like solutions near $\nu=1$ becomes less regular.
As the rotation rate increases, the last two effects become stronger.
This can be seen in Fig.~\ref{fig:spec_nu_04} for $\Omega/\Omega_{\rm K}=0.4$.
In particular, the two previously mentioned Rossby-like solutions almost behave as gravity-like solutions for $\nu>1$.

We focus now on how the modified Hough function $w_{\nu km}$ varies with the pseudo-radius $a$ and the horizontal coordinate $x$.
This dependence is illustrated for $m=-2$ and $\nu=2$ in Fig.~\ref{fig:func_nu2} at two different rotation rates.
Again, the solution at $a=0$ is not perturbed by the centrifugal acceleration, and is thus the same for all rotation rates.
As the pseudo-radius increases, Rossby-like solutions are slightly modified, whereas gravity-like solutions drastically change.
However, both types of solution seem to experience an avoided crossing, which is visible through the change in the number of nodes.
This behavior is qualitatively the same for the two rotation rates studied here.
Quantitatively, the dispersion of the eigenfunctions as a function of the pseudo-radius is larger for larger rotation rates.
The same eigenfunctions are plotted in Fig.~\ref{fig:func_nu10} for $m=-2$ and $\nu=10$.
Again, gravity-like solutions clearly experience an avoided crossing, but this is no longer visible for Rossby-like solutions.

\begin{figure*}
    \centering
    \resizebox{0.49\hsize}{!}{\includegraphics{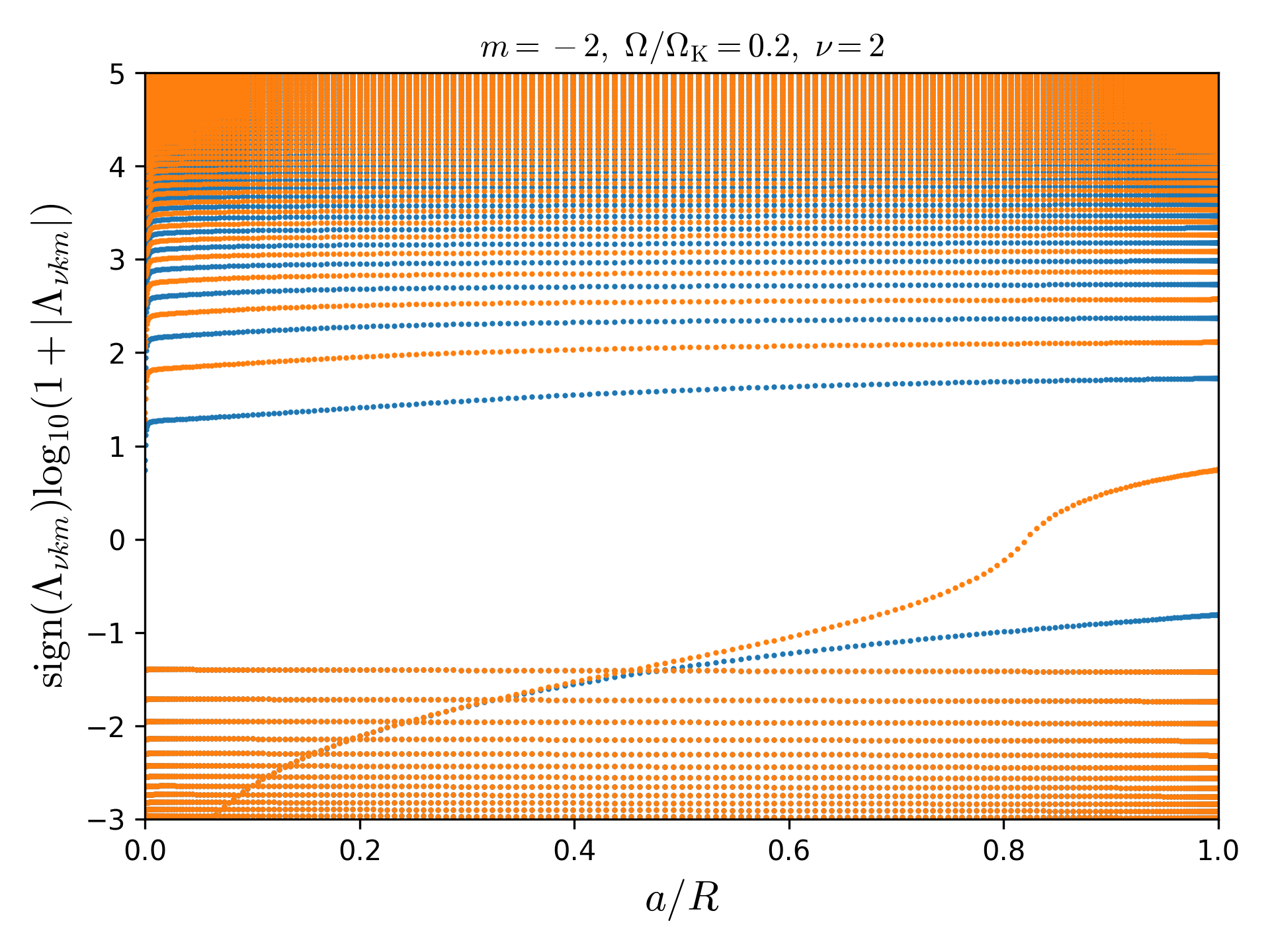}}
    \hfill
    \resizebox{0.49\hsize}{!}{\includegraphics{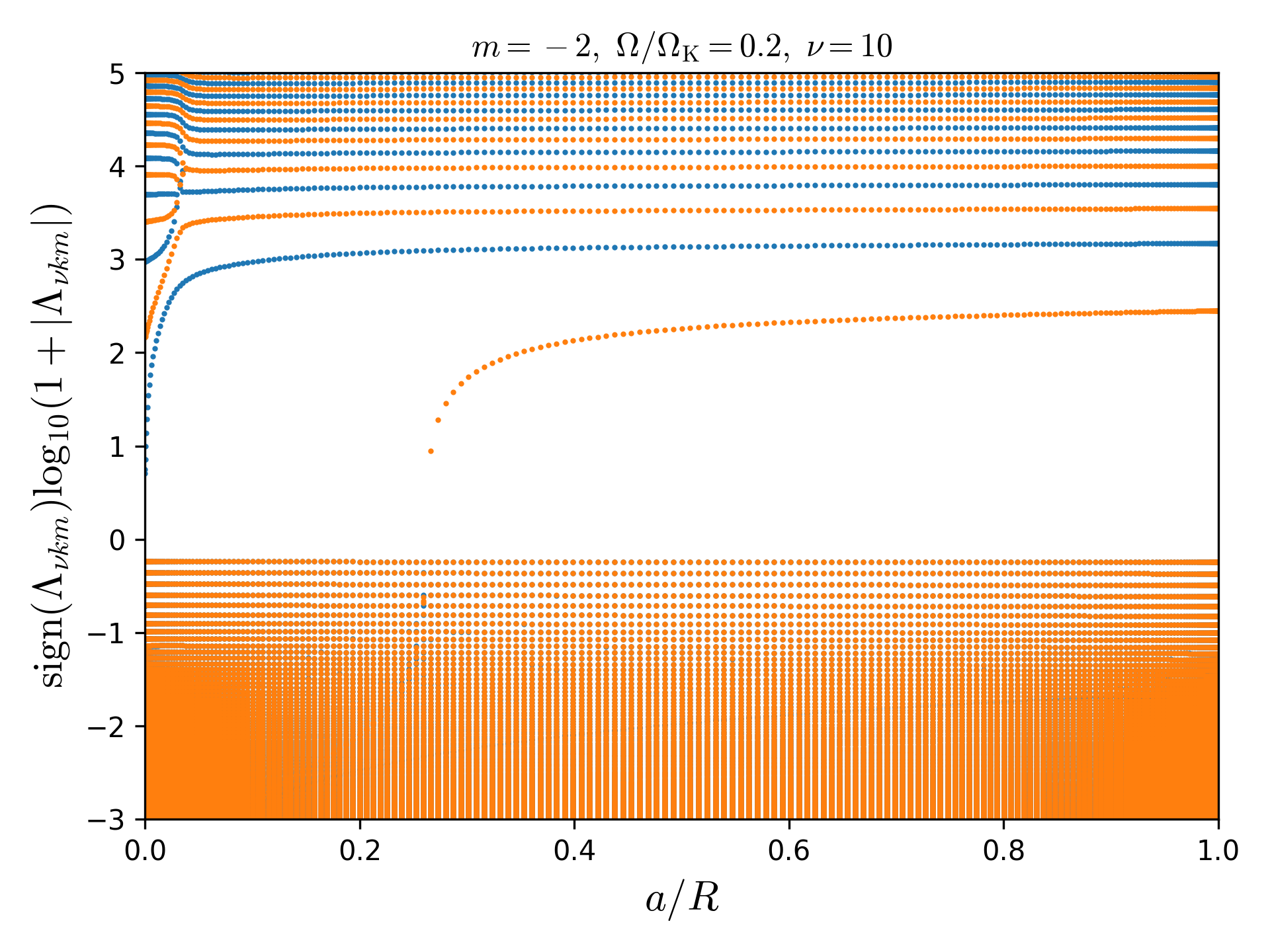}}
    \caption{Spectrum of the generalised Laplace tidal equation as a function of the pseudo-radius at $\Omega=0.2\Omega_{\rm K}$ and $m=-2$ for $\nu=2$ (left) and $\nu=10$ (right).
Blue (respectively orange) dots correspond to even (respectively odd) eigenfunctions.}
    \label{fig:spec_r_2}
\end{figure*}

The avoided crossings can be highlighted by plotting the eigenvalues as a function of the pseudo-radius (see Fig.~\ref{fig:spec_r_2}).
For $m=-2$ and $\nu=2$, the avoided crossing of Rossby-like solutions is clearly visible.
In contrast, that of gravity-like solutions is not because it occurs at very small pseudo-radii.
For $m=-2$ and $\nu=10$, it occurs at larger pseudo-radii, and is thus visible.

We finally compute the integral
\begin{equation}
    \label{eq:int}
    I = \int_{a_{\rm t1}}^{a_{\rm t2}}\frac{\Lambda^{1/2}_{\nu k m}\left(a\right)N\left(a\right)}{a}{\rm d}a
\end{equation}
at a fixed spin factor for different rotation rates to investigate how mode frequencies could be affected by the centrifugal acceleration.
The obtaind values are shown in Table~\ref{tab:int}.
\begin{table}
    \centering
    \(
    \begin{array}{c|cccc}
    \hline
    \hline
        \Omega/\Omega_{\rm K}   &   0.1     &   0.2     &   0.3     &   0.4    \\\hline
        I ({\rm s}^{-1})        &   0.0864  &   0.1024  &   0.1083  &   0.1133 \\
    \hline
    \end{array}
\)
\caption{Values of the integral $I$ (Eq.~\ref{eq:int}) as a function of the rotation rate for the gravity-like eigenfunctions plotted in Fig.~\ref{fig:func_nu10} ($m=-2$, $k=0$, $\nu=10$).}
    \label{tab:int}
\end{table}
The value of the integral slightly increases with the rotation rate, which will have an effect on the quantisation condition Eq.~\eqref{eq:quant}, and thus on the mode frequencies.

In addition to these frequency shifts, the changes in the wave structure induced by the centrifugal deformation observed in this section may also have an impact on the transport of angular momentum and on the tidal dissipation they induce \citep[e.g.][]{Mathis2009,BravinerOgilvie2015}. 

\section{Conclusion}
\label{sec:CP}

In this theoretical article, we generalize the Traditional Approximation of Rotation to the case where the deformation of a star (or planet) by the centrifugal acceleration is taken into account. We identify that the mathematical complexity introduced by the centrifugal acceleration is very similar to the one appearing when applying the TAR to differentially rotating spherical stars \citep{OgilvieLin2004,Mathis2009,VanReethetal2018}. Combining the TAR with the anelastic and the JWKB approximations, we derive a generalised Tidal Laplace Equation, which is a second-order linear Ordinary Differential Equation in $x=\cos\theta$ ($\theta$ being the colatitude) only with a parametric dependence on $a$ of its coefficients. The problem thus reduces to a classical Sturm-Liouville problem as in the case of uniformly rotating spherical stars. It allows us to derive the asymptotic frequencies of low-frequency gravito-inertial waves and the corresponding periods and period spacings. They can be used as a seismic probe of stellar interiors and rotation in moderately rapidly rotating deformed stars. In addition, the derived formalism can be used to study the angular momentum transport and tidal dissipation induced by low-frequency gravito-inertial waves in stars and planets. We have done a first numerical exploration of the eigenvalues and horizontal eigenfunctions of the generalised Laplace tidal equation following the methodology presented in Fig. \ref{Fig1}. We find that both gravity- and Rossby-like waves' eigenfunctions are affected by the centrifugal acceleration and vary with the pseudo-radii with a stronger deformation of the gravity-like solutions when compared to the spherical case. In this context, we see that both type of solutions are affected by avoided crossing phenomena. The next step will be to implement our equations in stellar evolution and oscillation codes \citep[as we did in the differentially rotating case in][]{VanReethetal2018} with comparing the obtained results with direct computations with 2D oscillation codes \citep{Reeseetal2006,Ballotetal2010,Ouazzanietal2012,Ouazzanietal2017} and to examine if the Traditional Approximation of Rotation can be generalised to the case of strongly deformed stars. 

\begin{acknowledgements}
We thank the referee for her/his constructive report which has allowed us to improve our work. S.~M. and V.~P. acknowledge support by ERC SPIRE 647383 and  by CNES PLATO \& GOLF grants at CEA-Saclay. They thank C. Aerts and C. Neiner for fruitful exchanges and K. Augustson for providing the $1.5M_{\odot}$ stellar model used in this work. S.~M. dedicates this work to the memory of his father in law and friend J. Neiner. 
\end{acknowledgements}


\bibliographystyle{aa}  
\bibliography{MathisTACAI}

\begin{thebibliography}{60}
\expandafter\ifx\csname natexlab\endcsname\relax\def\natexlab#1{#1}\fi

\bibitem[{{Aerts} {et~al.}(2010){Aerts}, {Christensen-Dalsgaard}, \&
  {Kurtz}}]{AJCDK2010}
{Aerts}, C., {Christensen-Dalsgaard}, J., \& {Kurtz}, D.~W. 2010,
  {Asteroseismology}

\bibitem[{{Aerts} {et~al.}(2018{\natexlab{a}}){Aerts}, {Mathis}, \&
  {Rogers}}]{Aertsetal2018b}
{Aerts}, C., {Mathis}, S., \& {Rogers}, T. 2018{\natexlab{a}}, arXiv e-prints
  [\eprint[arXiv]{1809.07779}]

\bibitem[{{Aerts} {et~al.}(2018{\natexlab{b}}){Aerts}, {Molenberghs},
  {Michielsen}, {Pedersen}, {Bj{\"o}rklund}, {Johnston}, {Mombarg}, {Bowman},
  {Buysschaert}, {P{\'a}pics}, {Sekaran}, {Sundqvist}, {Tkachenko}, {Truyaert},
  {Van Reeth}, \& {Vermeyen}}]{Aertsetal2018a}
{Aerts}, C., {Molenberghs}, G., {Michielsen}, M., {et~al.} 2018{\natexlab{b}},
  \apjs, 237, 15

\bibitem[{{Aerts} {et~al.}(2017){Aerts}, {Van Reeth}, \&
  {Tkachenko}}]{Aertsetal2017}
{Aerts}, C., {Van Reeth}, T., \& {Tkachenko}, A. 2017, \apjl, 847, L7

\bibitem[{{Alvan} {et~al.}(2015){Alvan}, {Strugarek}, {Brun}, {Mathis}, \&
  {Garcia}}]{Alvanetal2015}
{Alvan}, L., {Strugarek}, A., {Brun}, A.~S., {Mathis}, S., \& {Garcia}, R.~A.
  2015, \aap, 581, A112

\bibitem[{{Ballot} {et~al.}(2010){Ballot}, {Ligni{\`e}res}, {Reese}, \&
  {Rieutord}}]{Ballotetal2010}
{Ballot}, J., {Ligni{\`e}res}, F., {Reese}, D.~R., \& {Rieutord}, M. 2010,
  \aap, 518, A30

\bibitem[{{Berthomieu} {et~al.}(1978){Berthomieu}, {Gonczi}, {Graff},
  {Provost}, \& {Rocca}}]{Berthomieuetal1978}
{Berthomieu}, G., {Gonczi}, G., {Graff}, P., {Provost}, J., \& {Rocca}, A.
  1978, \aap, 70, 597

\bibitem[{{Bouabid} {et~al.}(2013){Bouabid}, {Dupret}, {Salmon},
  {Montalb{\'a}n}, {Miglio}, \& {Noels}}]{Bouabidetal2013}
{Bouabid}, M.-P., {Dupret}, M.-A., {Salmon}, S., {et~al.} 2013, \mnras, 429,
  2500

\bibitem[{{Braviner} \& {Ogilvie}(2014)}]{Bravineretal2014}
{Braviner}, H.~J. \& {Ogilvie}, G.~I. 2014, \mnras, 441, 2321

\bibitem[{{Braviner} \& {Ogilvie}(2015)}]{BravinerOgilvie2015}
{Braviner}, H.~J. \& {Ogilvie}, G.~I. 2015, \mnras, 447, 1141

\bibitem[{{Chandrasekhar}(1933)}]{Chandrasekhar1933}
{Chandrasekhar}, S. 1933, \mnras, 93, 390

\bibitem[{{Christophe} {et~al.}(2018){Christophe}, {Ballot}, {Ouazzani},
  {Antoci}, \& {Salmon}}]{Christopheetal2018}
{Christophe}, S., {Ballot}, J., {Ouazzani}, R.-M., {Antoci}, V., \& {Salmon},
  S.~J.~A.~J. 2018, \aap, 618, A47

\bibitem[{{Cohen-Tannoudji} {et~al.}(1986){Cohen-Tannoudji}, {Diu}, \&
  {Laloe}}]{CohenTannoudjietal1986}
{Cohen-Tannoudji}, C., {Diu}, B., \& {Laloe}, F. 1986, {Quantum Mechanics,
  Volume 2}, 626

\bibitem[{{Cowling}(1941)}]{Cowling1941}
{Cowling}, T.~G. 1941, \mnras, 101, 367

\bibitem[{{Dintrans} {et~al.}(1999){Dintrans}, {Rieutord}, \&
  {Valdettaro}}]{DintransRieutord1999}
{Dintrans}, B., {Rieutord}, M., \& {Valdettaro}, L. 1999, Journal of Fluid
  Mechanics, 398, 271

\bibitem[{{Fr{\"o}man} \& {Fr{\"o}man}(2005)}]{FromanFroman2005}
{Fr{\"o}man}, N. \& {Fr{\"o}man}, P.~O. 2005, {Physical Problems Solved by the
  Phase-Integral Method}, 228

\bibitem[{{Fuller} {et~al.}(2016){Fuller}, {Luan}, \&
  {Quataert}}]{Fulleretal2016}
{Fuller}, J., {Luan}, J., \& {Quataert}, E. 2016, \mnras, 458, 3867

\bibitem[{{Gerkema} \& {Shrira}(2005)}]{GerkemaShrira2005}
{Gerkema}, T. \& {Shrira}, V.~I. 2005, Journal of Fluid Mechanics, 529, 195

\bibitem[{{Gerkema} {et~al.}(2008){Gerkema}, {Zimmerman}, {Maas}, \& {van
  Haren}}]{Gerkemaetal2008}
{Gerkema}, T., {Zimmerman}, J.~T.~F., {Maas}, L.~R.~M., \& {van Haren}, H.
  2008, Reviews of Geophysics, 46, RG2004

\bibitem[{{Hough}(1898)}]{Hough1898}
{Hough}, S.~S. 1898, Philosophical Transactions of the Royal Society of London
  Series A, 191, 139

\bibitem[{{Kurtz} {et~al.}(2014){Kurtz}, {Saio}, {Takata}, {Shibahashi},
  {Murphy}, \& {Sekii}}]{Kurtzetal2014}
{Kurtz}, D.~W., {Saio}, H., {Takata}, M., {et~al.} 2014, \mnras, 444, 102

\bibitem[{{Lee}(1993)}]{Lee1993}
{Lee}, U. 1993, The Astrophysical Journal, 405, 359

\bibitem[{{Lee} \& {Baraffe}(1995)}]{LeeBaraffe1995}
{Lee}, U. \& {Baraffe}, I. 1995, Astronomy \& Astrophysics, 301, 419

\bibitem[{{Lee} {et~al.}(2014){Lee}, {Neiner}, \& {Mathis}}]{Leeetal2014}
{Lee}, U., {Neiner}, C., \& {Mathis}, S. 2014, \mnras, 443, 1515

\bibitem[{{Lee} \& {Saio}(1987)}]{LeeSaio1987}
{Lee}, U. \& {Saio}, H. 1987, \mnras, 224, 513

\bibitem[{{Lee} \& {Saio}(1989)}]{LeeSaio1989}
{Lee}, U. \& {Saio}, H. 1989, \mnras, 237, 875

\bibitem[{{Lee} \& {Saio}(1993)}]{LeeSaio1993}
{Lee}, U. \& {Saio}, H. 1993, \mnras, 261, 415

\bibitem[{{Lee} \& {Saio}(1997)}]{LeeSaio1997}
{Lee}, U. \& {Saio}, H. 1997, \apj, 491, 839

\bibitem[{{Li} {et~al.}(2019){Li}, {Bedding}, {Murphy}, {Van Reeth}, {Antoci},
  \& {Ouazzani}}]{Lietal2019}
{Li}, G., {Bedding}, T.~R., {Murphy}, S.~J., {et~al.} 2019, \mnras, 482, 1757

\bibitem[{{Longuet-Higgins}(1968)}]{LonguetHiggins1968}
{Longuet-Higgins}, M.~S. 1968, Philosophical Transactions of the Royal Society
  of London Series A, 262, 511

\bibitem[{{Mathis}(2009)}]{Mathis2009}
{Mathis}, S. 2009, \aap, 506, 811

\bibitem[{{Mathis} {et~al.}(2008){Mathis}, {Talon}, {Pantillon}, \&
  {Zahn}}]{Mathisetal2008}
{Mathis}, S., {Talon}, S., {Pantillon}, F.-P., \& {Zahn}, J.-P. 2008, \solphys,
  251, 101

\bibitem[{{Mathis} \& {Zahn}(2004)}]{MathisZahn2004}
{Mathis}, S. \& {Zahn}, J.-P. 2004, \aap, 425, 229

\bibitem[{{Murphy} {et~al.}(2016){Murphy}, {Fossati}, {Bedding}, {Saio},
  {Kurtz}, {Grassitelli}, \& {Wang}}]{Murphyetal2016}
{Murphy}, S.~J., {Fossati}, L., {Bedding}, T.~R., {et~al.} 2016, \mnras, 459,
  1201

\bibitem[{{Ogilvie} \& {Lin}(2004)}]{OgilvieLin2004}
{Ogilvie}, G.~I. \& {Lin}, D.~N.~C. 2004, \apj, 610, 477

\bibitem[{{Ogilvie} \& {Lin}(2007)}]{OgilvieLin2007}
{Ogilvie}, G.~I. \& {Lin}, D.~N.~C. 2007, \apj, 661, 1180

\bibitem[{{Ouazzani} {et~al.}(2012){Ouazzani}, {Dupret}, \&
  {Reese}}]{Ouazzanietal2012}
{Ouazzani}, R.-M., {Dupret}, M.-A., \& {Reese}, D.~R. 2012, \aap, 547, A75

\bibitem[{{Ouazzani} {et~al.}(2019){Ouazzani}, {Marques}, {Goupil},
  {Christophe}, {Antoci}, {Salmon}, \& {Ballot}}]{Ouazzanietal2018}
{Ouazzani}, R.~M., {Marques}, J.~P., {Goupil}, M.~J., {et~al.} 2019, \aap, 626,
  A121

\bibitem[{{Ouazzani} {et~al.}(2017){Ouazzani}, {Salmon}, {Antoci}, {Bedding},
  {Murphy}, \& {Roxburgh}}]{Ouazzanietal2017}
{Ouazzani}, R.-M., {Salmon}, S.~J.~A.~J., {Antoci}, V., {et~al.} 2017, \mnras,
  465, 2294

\bibitem[{{Paxton} {et~al.}(2018){Paxton}, {Schwab}, {Bauer}, {Bildsten},
  {Blinnikov}, {Duffell}, {Farmer}, {Goldberg}, {Marchant}, {Sorokina},
  {Thoul}, {Townsend}, \& {Timmes}}]{Paxtonetal2018}
{Paxton}, B., {Schwab}, J., {Bauer}, E.~B., {et~al.} 2018, \apjs, 234, 34

\bibitem[{{Pedersen} {et~al.}(2018){Pedersen}, {Aerts}, {P{\'a}pics}, \&
  {Rogers}}]{Pedersenetal2018}
{Pedersen}, M.~G., {Aerts}, C., {P{\'a}pics}, P.~I., \& {Rogers}, T.~M. 2018,
  \aap, 614, A128

\bibitem[{{Prat} {et~al.}(2016){Prat}, {Ligni{\`e}res}, \&
  {Ballot}}]{Pratetal2016}
{Prat}, V., {Ligni{\`e}res}, F., \& {Ballot}, J. 2016, \aap, 587, A110

\bibitem[{{Prat} {et~al.}(2018){Prat}, {Mathis}, {Augustson}, {Ligni{\`e}res},
  {Ballot}, {Alvan}, \& {Brun}}]{Pratetal2018}
{Prat}, V., {Mathis}, S., {Augustson}, K., {et~al.} 2018, \aap, 615, A106

\bibitem[{{Prat} {et~al.}(2017){Prat}, {Mathis}, {Ligni{\`e}res}, {Ballot}, \&
  {Culpin}}]{Pratetal2017}
{Prat}, V., {Mathis}, S., {Ligni{\`e}res}, F., {Ballot}, J., \& {Culpin}, P.-M.
  2017, \aap, 598, A105

\bibitem[{{Reese} {et~al.}(2006){Reese}, {Ligni{\`e}res}, \&
  {Rieutord}}]{Reeseetal2006}
{Reese}, D., {Ligni{\`e}res}, F., \& {Rieutord}, M. 2006, \aap, 455, 621

\bibitem[{{Roxburgh}(2006)}]{Roxburgh2006}
{Roxburgh}, I.~W. 2006, \aap, 454, 883

\bibitem[{{Saio}(1981)}]{Saio1981}
{Saio}, H. 1981, The Astrophysical Journal, 244, 299

\bibitem[{{Saio} {et~al.}(2015){Saio}, {Kurtz}, {Takata}, {Shibahashi},
  {Murphy}, {Sekii}, \& {Bedding}}]{Saioetal2015}
{Saio}, H., {Kurtz}, D.~W., {Takata}, M., {et~al.} 2015, \mnras, 447, 3264

\bibitem[{{Smeyers} \& {Denis}(1971)}]{SmeyersDenis1971}
{Smeyers}, P. \& {Denis}, J. 1971, Astronomy \& Astrophysics, 14, 311

\bibitem[{{Sweet}(1950)}]{Sweet1950}
{Sweet}, P.~A. 1950, \mnras, 110, 548

\bibitem[{{Tassoul}(1978)}]{Tassoul1978}
{Tassoul}, J.-L. 1978, {Theory of rotating stars}

\bibitem[{{Tassoul}(1980)}]{Tassoul1980}
{Tassoul}, M. 1980, \apjs, 43, 469

\bibitem[{{Townsend}(2003)}]{Townsend2003}
{Townsend}, R.~H.~D. 2003, \mnras, 340, 1020

\bibitem[{{Van Reeth} {et~al.}(2018){Van Reeth}, {Mombarg}, {Mathis},
  {Tkachenko}, {Fuller}, {Bowman}, {Buysschaert}, {Johnston}, {Garc{\'{\i}}a
  Hern{\'a}ndez}, {Goldstein}, {Townsend}, \& {Aerts}}]{VanReethetal2018}
{Van Reeth}, T., {Mombarg}, J.~S.~G., {Mathis}, S., {et~al.} 2018, \aap, 618,
  A24

\bibitem[{{Van Reeth} {et~al.}(2016){Van Reeth}, {Tkachenko}, \&
  {Aerts}}]{VanReethetal2016}
{Van Reeth}, T., {Tkachenko}, A., \& {Aerts}, C. 2016, \aap, 593, A120

\bibitem[{{Van Reeth} {et~al.}(2015{\natexlab{a}}){Van Reeth}, {Tkachenko},
  {Aerts}, {P{\'a}pics}, {Degroote}, {Debosscher}, {Zwintz}, {Bloemen}, {De
  Smedt}, {Hrudkova}, {Raskin}, \& {Van Winckel}}]{VanReethetal2015a}
{Van Reeth}, T., {Tkachenko}, A., {Aerts}, C., {et~al.} 2015{\natexlab{a}},
  \aap, 574, A17

\bibitem[{{Van Reeth} {et~al.}(2015{\natexlab{b}}){Van Reeth}, {Tkachenko},
  {Aerts}, {P{\'a}pics}, {Triana}, {Zwintz}, {Degroote}, {Debosscher},
  {Bloemen}, {Schmid}, {De Smedt}, {Fremat}, {Fuentes}, {Homan}, {Hrudkova},
  {Karjalainen}, {Lombaert}, {Nemeth}, {{\O}stensen}, {Van De Steene}, {Vos},
  {Raskin}, \& {Van Winckel}}]{VanReethetal2015b}
{Van Reeth}, T., {Tkachenko}, A., {Aerts}, C., {et~al.} 2015{\natexlab{b}},
  \apjs, 218, 27

\bibitem[{{Wang} {et~al.}(2016){Wang}, {Boyd}, \& {Akmaev}}]{Wangetal2016}
{Wang}, H., {Boyd}, J.~P., \& {Akmaev}, R.~A. 2016, Geoscientific Model
  Development, 9, 1477

\bibitem[{{Zahn}(1966)}]{Zahn1966a}
{Zahn}, J.~P. 1966, Annales d'Astrophysique, 29, 313

\bibitem[{{Zahn}(1992)}]{Zahn1992}
{Zahn}, J.-P. 1992, \aap, 265, 115

\end{thebibliography}



\begin{appendix}

\section{The deformation of a moderately rotating star}
\label{Appendix:deformation}

The objective of this appendix is to determine the deformation of an isobar in the case of a moderately and uniformly rotating star where the centrifugal acceleration is a linear perturbation of the order of $\left(\Omega/\Omega_{\rm K}\right)^2\equiv\left(\Omega/\sqrt{GM/R^3}\right)^2$ ($\Omega_{\rm K}$ is the Keplerian critical angular velocity, $M$ and $R$ are the stellar mass and radius, respectively, and $G$ is the universal constant of gravity). 

The first step is to calculate the perturbation of the gravitational potential $\phi$ on the sphere of radius $r$. We expand $\phi$ on the orthogonal basis formed by the Legendre polynomials:  
\begin{equation}
\phi(r, \theta)=\phi_{0}(r)+\phi_{1}(r,\theta)=\phi_{0}(r)+\sum_{l}{\phi}_{l}(r)P_{l}(\cos\theta), 
\end{equation} 
where $\phi_{0}=-GM_r(r)/r$ is the gravitational potential of the non-rotating star ($M_r$ being the mass inside the sphere of radius $r$) and $\phi_{1}$ its perturbation induced by the centrifugal acceleration. 
We follow the method of linearisation of the hydrostatic balance developed in \cite{Sweet1950}, \cite{Zahn1966a,Zahn1992}, and \cite{MathisZahn2004}, and expand the pressure and the density around the sphere in the same way that $\phi$:   
\begin{eqnarray}
P(r, \theta)&\!=\!&P_{0}(r)+P_{1}(r,\theta)\!=\!P_{0}(r)+\sum_{l}{P}_{l}(r)P_{l}(\cos\theta),\\
\rho(r, \theta)&\!=\!&\rho_{0}(r)+\rho_{1}(r,\theta)\!=\!\rho_{0}(r)+\sum_{l}{\rho}_{l}(r)P_{l}(\cos\theta).
\end{eqnarray}
Then, we take the hydrostatic equation   
\begin{equation}
\frac{\boldsymbol\nabla P}{\rho}= - \boldsymbol\nabla\phi+\boldsymbol{\mathcal F}_{\rm C}, \quad \hbox{ where } \quad 
\boldsymbol{\mathcal F}_{\rm C} =\frac{1}{2}\Omega^{2}\vec\nabla(r^2\sin^2\theta)
\end{equation} 
is the centrifugal acceleration, which derives from a potential: 
\begin{equation}
\boldsymbol{\mathcal F}_{\rm C}=-\boldsymbol\nabla U \quad \hbox{ with } \quad U=-\frac{1}{2}\Omega^{2}r^2\sin^2\theta
\end{equation}
in the case considered here of a uniform rotation.
The hydrostatic balance thus becomes
\begin{equation}
\boldsymbol\nabla P=-\rho \boldsymbol\nabla\left(\phi+U\right),
\label{eq:hydrocentrifuge}
\end{equation}
which we expand to the first order as
\begin{equation}
\boldsymbol\nabla P_{1}=-\rho_0\boldsymbol\nabla\left(\phi_1+U\right)-\rho_1\boldsymbol\nabla\phi_{0}.
\end{equation}
Taking the curl of Eq. (\ref{eq:hydrocentrifuge}), we also have:
\begin{equation}
\boldsymbol\nabla\rho\times\boldsymbol\nabla\left(\phi+U\right)=0.
\end{equation}
The equipotential for $\left(\phi+U\right)$, the isodensity and the isobar thus coincide. As a consequence $P$ can be written as a function of $\left(\phi+U\right)$:
\begin{equation}
P={\mathcal F}\left(\phi+U\right).
\end{equation}
When linearised to the first-order, we get
\begin{equation}
P_{1}=\frac{{\rm d}{\mathcal F}}{{\rm d}\phi_0}\left(\phi_1+U\right)=-\rho_0\left(\phi_1+U\right)
\label{eq:perturblat}
\end{equation}
since ${\rm d}{\mathcal F}/{\rm d}\phi_0={\rm d}P_0/{\rm d}\phi_0=-\rho_0$. This leads to
\begin{equation}
{\boldsymbol\nabla}P_1=-\rho_0{\boldsymbol\nabla}\left(\phi_1+U\right)-{\boldsymbol\nabla}\rho_0\left(\phi_1+U\right)
\end{equation}
that provides us the perturbation of density
\begin{equation}
\rho_1=\frac{1}{g_0\left(r\right)}\frac{{\rm d}\rho_0}{{\rm d}r}\left(\phi_1+U\right),
\end{equation}
where $g_{0}=GM_r(r)/r^2$. Next, we insert the modal expansion  of $\rho_{1}$ and those of the centrifugal potential:
\begin{eqnarray}
U=\sum_l U_l\left(r\right)P_l\left(\cos\theta\right),\quad\hbox{where}\quad l=\left\{0,2\right\}
\label{ab}
\end{eqnarray}
and
\begin{eqnarray}
U_0&=&-\frac{1}{3}\,\Omega^2r^2\\
U_2&=&\frac{1}{3}\,\Omega^2r^2.
\end{eqnarray}
This yields the modal amplitude  of the density fluctuation over the sphere
\begin{equation}
{\rho}_{l}(r) = 
\frac{1}{g_{0}}\frac{{\rm d}\rho_{0}}{{\rm d}r}\left({\phi}_{l}+U_{l}\right).
\end{equation}  
We insert this expression in the perturbed Poisson equation $\nabla^2 {\phi}_{l} =
 4 \pi G {\rho}_{l}$ an we retrieve the \cite{Sweet1950} and \cite {Zahn1966a} result
 \begin{equation}
 \frac{1}{r}\frac{{\rm d}^2}{{\rm d}r^2}\left(r{\phi}_{l}\right)-\frac{l(l+1)}{r^2}{\phi}_{l}-\frac{4\pi G}{g_{0}}\frac{{\rm d}\rho_{0}}{{\rm d}r}{\phi}_{l}=\frac{4\pi G}{g_{0}}\frac{{\rm d}\rho_{0}}{{\rm d}r}U_{l}
 \label{poisson-pert}
 \end{equation}
with $l=\left\{0,2\right\}$. The applied boundary conditions are
\begin{equation}
{\phi}_{l}=0\quad\hbox{at}\quad r=0\quad\hbox{and}\quad\frac{\rm d}{{\rm d}r}{\phi}_{l}-\frac{\left(l+1\right)}{r}{\phi}_{l}=0\quad\hbox{at}\quad r=R,
\label{bc}
\end{equation}
$R$ being the star's (or the planet's) surface radius.

Taking the latitudinal component of the hydrostatic balance (Eq. \ref{eq:perturblat}) finally provides us the radial functions of the pressure fluctuation expansion
 \begin{equation}
{P}_{l}= -\rho_{0}\left({\phi}_{l}+U_{l}\right).
\label{eq:HydrostatTheta}
 \end{equation}\\
  
We introduce the radial coordinate of the isobar
\begin{equation}
a_P(r, \theta) = r + \sum_{l}\xi_{l}(r) P_{l}(\cos\theta).
\end{equation}
Taking the Taylor expansion of $P$ to first order, we have:  
\begin{eqnarray}  
\lefteqn{P\left(r+\sum_{l}\xi_{l}(r)P_{l}(\cos\theta),\theta\right)=P_{0}(r)}\nonumber\\
&+&\sum_{l}{P}_{l}(r)P_{l}(\cos\theta)+ \left(\frac{ {\rm d} P_{0}}{ {{\rm d} r}}\right)   \sum_{l}\xi_{l}(r)P_{l}(\cos\theta).
\end{eqnarray}
By definition the pressure is constant on the isobar. We conclude that  
\begin{equation}
\xi_{l}(r)=-\frac{{P}_{l}} {{{\rm d} P_{0}/{\rm d}r}}=-\frac{\left({\phi}_{l}+U_{l}\right)}{g_0},
\end{equation}
where we have used Eq. (\ref{eq:HydrostatTheta}) and the zeroth-order hydrostatic balance ${{\rm d} P_{0}/{\rm d}r}=-\rho_{0}g_{0}$.\\

We finally introduce the pseudo-radial coordinate $a$ defined in Eq.~\eqref{isopot}.
\begin{equation}
r=a\left[1+\varepsilon\left(a,\theta\right)\right],
\end{equation}
where $\varepsilon$ is also expanded on Legendre polynomials as
\begin{equation}
\varepsilon\left(a,\theta\right)=\sum_{l}{\varepsilon}_{l}\left(a\right)P_{l}\left(\cos\theta\right).
\label{epsilon}
\end{equation} 
In contrast to \citet{Lee1993}, here we do not use the Chandrasekhar-Milne expansion \citep[e.g.][]{Chandrasekhar1933,Tassoul1978} to compute $a$, because it leads to infinite $\varepsilon$ at the center.
Instead, we use a simple mapping such that $r$ is equal to the deformed stellar radius $r_{\rm s}(\theta)$ when $a=R$ and that $r\simeq a$ near the center.
The simplest mapping verifying these conditions is
\begin{equation}
    r = a + \left(\frac{r_{\rm s}}{R}-1\right)\frac{a^2}{R}.
\end{equation}
At first order, this leads to
\begin{equation}
    \varepsilon_{l}=\frac{\xi_l(R)a}{R^2}=-\frac{{\phi}_{l}(R)+U_l(R)}{g_0(R)}\frac{a}{R^2}\quad\hbox{with}\quad l=\left\{0,2\right\}.
\label{shape}
\end{equation}

\section{Perturbative analytical solutions}
\label{Appendix:perturbation}

\subsection{Propagative waves with fixed frequencies}
\label{sec:fixed}

We consider here any propagative waves with non-quantized frequencies ($\omega$) and thus spin parameter ($\nu=2\Omega/\omega$). This is for instance the case of tidally-excited waves \citep[e.g.][]{OgilvieLin2004,OgilvieLin2007,Bravineretal2014} where the tidal frequency is fixed by the difference between the angular velocities of the primary and of the orbit of the companion, and of progressive waves \citep[e.g.][]{Alvanetal2015}.

As described above in Appendix \ref{Appendix:deformation}, the structure of a moderatly rotating body where the centifugal acceleration can be treated as a linear perturbation is the linear combination of the non-rotating structure and of a perturbation ($\varepsilon$) of the order of $\left(\Omega/\sqrt{GM/R^3}\right)^2$. Therefore, we can make a corresponding linear expansion in $\varepsilon$ of the generalised Laplace tidal operator (${\mathcal L}_{\nu k m}$) derived in Eq. (\ref{tidal}) and of its eigenvalue ($\Lambda_{\nu k m}$) using the linear perturbation theory as in quantum mechanics \citep[][]{CohenTannoudjietal1986}. We obtain
\begin{equation}
{\mathcal L}_{\nu m}={\mathcal L}_{\nu m}^{\left(0\right)}+{\mathcal L}_{\nu m}^{\left(1\right)},
\end{equation}
where
\begin{equation}
{\mathcal L}_{\nu m}^{\left(0\right)}=\frac{{\rm d}}{{\rm d}x}\left(\frac{1-x^2}{1-\nu^2x^2}\frac{\rm d}{{\rm d}x}\right)-\frac{1}{1-\nu^2x^2}\left(\frac{m^2}{1-x^2}+m\nu\frac{1+\nu^2 x^2}{1-\nu^2 x^2}\right)
\end{equation}
with
\begin{equation}
{\mathcal L}_{\nu m}^{\left(0\right)}\left[\Theta_{\nu km}\left(x\right)\right]=-\Lambda_{\nu km}^{\left(0\right)}\Theta_{\nu km}\left(x\right)
\end{equation}
is the usual Laplace tidal operator in the spherical case with its Hough eigenfunctions ($\Theta_{\nu km}$) and eigenvalues ($\Lambda_{\nu km}^{\left(0\right)}$), and ${\mathcal L}_{\nu m}^{\left(1\right)}$ is its first-order centrifugal correction
\begin{equation}
{\mathcal L}_{\nu m}^{\left(1\right)}=\partial_{x}\left[C_{1}\left(a,x\right)\partial_{x}\right]+C_{2}\left(a,x\right)\partial_{x}+C_{3}\left(a,x\right)
 \label{eq:PLTO}
\end{equation}
with
\begin{equation}
C_{1}\left(a,x\right)=\frac{-2\left(1-x^2\right)}{\left(1-\nu^2 x^2\right)^2}\left[\left(1-\nu^2x^2\right){\varepsilon}+x\left(1-x^2\right)\nu^2\partial_x{\varepsilon}\right],\nonumber\\
\end{equation}
\begin{equation}
C_{2}\left(a,x\right)=\frac{\left(1-x^2\right)}{\left(1-\nu^2 x^2\right)}\left(3\partial_{x}\varepsilon+a\partial_{a,x}\varepsilon\right),
\end{equation}
and
\begin{eqnarray}
\lefteqn{C_{3}\left(a,x\right)=\frac{2m^2\left[\left(1-\nu^2x^2\right)\varepsilon+x\nu^2\left(1-x^2\right)\partial_{x}\varepsilon\right]}{\left(1-x^2\right)\left(1-\nu^2x^2\right)^2}}\nonumber\\
&+&\frac{m\nu}{\left(1-\nu^2x^2\right)^3}\left[2\left(1-\nu^4x^4\right)\varepsilon\right.\nonumber\\
&&+2\nu^2x\left(3+x^2\left[\left(1+x^2\right)\nu^2-5\right]\right)\partial_{x}\varepsilon \nonumber\\
&&\left.+\left(1-x^2\right)\left(1-\nu^4x^4\right)\partial_{x,x}\varepsilon\right]\nonumber\\
&-&\frac{m\nu x\left(3\partial_{x}\varepsilon+a\partial_{a,x}\varepsilon\right)}{1-\nu^2 x^2}.
\end{eqnarray}
These coefficients are obtained using the first-order expansion of ${\mathcal C}$ and ${\mathcal D}$ given in Tab.\ref{tab:coefficient} that become
\begin{eqnarray}
{\mathcal C}&=&1-\frac{1-x^2}{x}\partial_x\varepsilon\\
{\mathcal D}&=&\left(1-\nu^2x^2\right)\left(1+2\varepsilon+\frac{2\nu^2 x \left(1-x^2\right)\partial_x\varepsilon}{1-\nu^2 x^2}\right),
\label{linCD}
\end{eqnarray}
while ${\mathcal A}$, ${\mathcal B}$, and ${\mathcal E}$ are unchanged.

The corresponding eigenvalues and eigenfunctions are derived
\begin{eqnarray}
\Lambda_{\nu km}\left(a\right)&=&\Lambda_{\nu km}^{\left(0\right)}+\Lambda_{\nu km}^{\left(1\right)}\left(a\right),\\
w_{\nu k m}\left(a,x\right)&=&\Theta_{\nu k m}\left(x\right)+w_{\nu k m}^{\left(1\right)}\left(a,x\right),
\end{eqnarray}
with their respective centrifugal perturbations
\begin{eqnarray}
\Lambda_{\nu km}^{\left(1\right)}\left(a\right)&=&\frac{\int_{-1}^{1}\Theta_{\nu km}\left(x\right){\mathcal L}_{\nu m}^{\left(1\right)}\left[\Theta_{\nu km}\left(x\right)\right]{\rm d}x}{\int_{-1}^{1}\left[\Theta_{\nu km}\left(x\right)\right]^2{\rm d}x},\\
w_{\nu k m}^{\left(1\right)}\left(a,x\right)&=&\sum_{k^{'}\ne k}\frac{\int_{-1}^{1}\Theta_{\nu km}\left(x\right){\mathcal L}_{\nu m}^{\left(1\right)}\left[\Theta_{\nu k'm}\left(x\right)\right]{\rm d}x}{\Lambda_{\nu km}^{\left(0\right)}-\Lambda_{\nu k'm}^{\left(0\right)}}\Theta_{\nu k^{'} m}\left(x\right).\nonumber\\
\end{eqnarray}
Using the polarisation relations (Eq. \ref{polarisation}), the same expansion is done for the JWKB amplitudes of the Lagrangian displacement:
\begin{equation}
{\widehat\xi}_{j;\nu km}\left(a,x\right)={\widehat\xi}_{j;\nu km}^{\,\,\left(0\right)}\left(a,x\right)+{\widehat\xi}_{j;\nu km}^{\,\,\left(1\right)}\left(a,x\right),
\end{equation}
where $j\equiv\left\{r,\theta,\varphi\right\}$.

We get in the vertical direction
\begin{eqnarray}
{\widehat\xi}_{r;\nu km}^{\,\,\left(0\right)}&=&-i\frac{k_{V;\nu k m}^{\left(0\right)}}{N^2}\Theta_{\nu km},\\
{\widehat\xi}_{r;\nu km}^{\,\,\left(1\right)}&=&-\frac{i}{N^2}\left(k_{V;\nu k m}^{\left(0\right)}w_{\nu k m}^{\left(1\right)}+k_{V;\nu k m}^{\left(1\right)}\Theta_{\nu km}\right),
\end{eqnarray}
where
\begin{eqnarray}
k_{V;\nu k m}^{\left(0\right)}&=&\frac{N\left(a\right)}{\omega_{k m}}\frac{\sqrt{\Lambda_{\nu k m}^{\left(0\right)}}}{a},\\
k_{V;\nu k m}^{\left(1\right)}&=&\frac{1}{2}\frac{N\left(a\right)}{\omega_{k m}}\frac{\sqrt{\Lambda_{\nu k m}^{\left(0\right)}}}{a} \frac{\Lambda_{\nu k m}^{\left(1\right)}\left(a\right)}{\Lambda_{\nu k m}^{\left(0\right)}}.
\end{eqnarray}
We obtain for the latitudinal and azimuthal directions
\begin{eqnarray}
{\widehat\xi}_{\theta;\nu km}^{\,\,\left(0\right)}&=&{\mathcal L}_{\nu m}^{\theta;\left(0\right)}\Theta_{\nu km},\\
{\widehat\xi}_{\theta;\nu km}^{\,\,\left(1\right)}&=&{\mathcal L}_{\nu m}^{\theta;\left(0\right)}w_{\nu k m}^{\left(1\right)}+{\mathcal L}_{\nu m}^{\theta;\left(1\right)}\Theta_{\nu km}
\end{eqnarray}
and
\begin{eqnarray}
{\widehat\xi}_{\varphi;\nu km}^{\,\,\left(0\right)}&=&{\mathcal L}_{\nu m}^{\varphi;\left(0\right)}\Theta_{\nu km},\\
{\widehat\xi}_{\varphi;\nu km}^{\,\,\left(1\right)}&=&{\mathcal L}_{\nu m}^{\varphi;\left(0\right)}w_{\nu k m}^{\left(1\right)}+{\mathcal L}_{\nu m}^{\varphi;\left(1\right)}\Theta_{\nu km},
\end{eqnarray}
where
\begin{eqnarray}
{\mathcal L}_{\nu m}^{\theta;\left(0\right)}&=&\frac{1}{a}\frac{1}{\omega^2}\frac{1}{\sqrt{1-x^2}}\left[-\left(1-x^2\right)C_{4}^{\left(0\right)}\partial_x+m\nu x C_{5}^{\left(0\right)}\right],\\
{\mathcal L}_{\nu m}^{\theta;\left(1\right)}&=&\frac{1}{a}\frac{1}{\omega^2}\frac{1}{\sqrt{1-x^2}}\left[-\left(1-x^2\right)C_{4}^{\left(1\right)}\partial_x+m\nu x C_{5}^{\left(1\right)}\right],\\
{\mathcal L}_{\nu m}^{\varphi;\left(0\right)}&=&\frac{i}{a}\frac{1}{\omega^2}\frac{1}{\sqrt{1-x^2}}\left[-\nu x \left(1-x^2\right)C_{5}^{\left(0\right)}\partial_x+m C_{4}^{\left(0\right)}\right],\\
{\mathcal L}_{\nu m}^{\varphi;\left(1\right)}&=&\frac{i}{a}\frac{1}{\omega^2}\frac{1}{\sqrt{1-x^2}}\left[-\nu x \left(1-x^2\right)C_{5}^{\left(1\right)}\partial_x+m C_{4}^{\left(1\right)}\right]
\end{eqnarray}
with
\begin{eqnarray}
C_{4}^{\left(0\right)}&=&C_{5}^{\left(0\right)}=\frac{1}{1-\nu^2 x^2},\\
C_{4}^{\left(1\right)}&=&\frac{-2\left[\left(1-\nu^2 x^2\right)\varepsilon+x\left(1-x^2\right)\nu^2\partial_{x}\varepsilon\right]}{\left(1-\nu^2 x^2\right)^2},\\
C_{5}^{\left(1\right)}&=&-\frac{2 x \left(1-\nu^2 x^2\right)\varepsilon+\left(1-x^2\right)\left(1+\nu^2 x^2\right)\partial_x\varepsilon}{x\left(1-\nu^2x^2\right)^2}.
\end{eqnarray}
This analytical solution using first-order perturbative method can be of great interest to study propagative low-frequency GIWs in moderately rotating stars \citep[their excitation, their propagation, their damping and the potential angular momentum transport they induce; e.g.][]{Mathis2009}. It can also be used to compute tidal dissipation in slightly deformed stably stratified stellar and planetary layers \citep[][]{Bravineretal2014} like in the case of Saturn for instance \citep[][]{Fulleretal2016}.\\

Finally, the first-order linear perturbation theory at fixed given frequency can be used to compute the modification of the critical colatitude $\theta_{\rm c}$ for which
\begin{equation}
\mathcal{D}(a,\theta_{\rm c})=0.
\end{equation}
Using the linearisation of ${\mathcal C}$ given in Eq. (\ref{linCD}), this becomes: 
\begin{equation}
1-\nu^2\cos^2\theta_{\rm c}[1+\tan\theta_{\rm c}\partial_\theta\varepsilon(a,\theta_{\rm c})]^2 = 0.
\end{equation}
Posing $\theta_{\rm c}=\theta_{\rm c}^{(0)}+\theta_{\rm c}^{(1)}$, where $\theta_{\rm c}^{(0)}=\arccos(1/\nu)$ is the classical critical colatitude \citep[e.g.][]{LeeSaio1997} and $\theta_{\rm c}^{(1)}$ scales with $\varepsilon$, we obtain at the first order in $\varepsilon$
\begin{equation}
1-\nu^2\cos^2\theta_{\rm c}^{(0)}\left\{1+2\tan\theta_{\rm c}^{(0)}\left[\partial_\theta\varepsilon(a,\theta_{\rm c}^{(0)}) - \theta_{\rm c}^{(1)}\right]\right\}=0
\end{equation}
that leads to
\begin{equation}
\theta_{\rm c}^{(1)}=\partial_\theta\varepsilon(a,\theta_{\rm c}^{(0)}).
\end{equation}
In the case of uniform rotation
\begin{equation}
\varepsilon(a,\theta) = \varepsilon_0(a) + \varepsilon_2(a)P_2(\cos\theta)
\end{equation}
and thus
\begin{equation}
\partial_\theta\varepsilon(a,\theta_{\rm c}^{(0)}) \simeq -3\varepsilon_2(a)\sin\theta_{\rm c}^{(0)}\cos\theta_{\rm c}^{(0)}.
\end{equation}
Using the expression for $\theta_{\rm c}^{(0)}$ as a function of $\nu$, the critical colatitude finally reads
\begin{equation}
\theta_{\rm c}\simeq\arccos(1/\nu)-\frac{3\varepsilon_2}{\nu}\sqrt{1-\frac{1}{\nu^2}}.
\label{CritLatCA}
\end{equation}
If one makes the rough assumption that $\varepsilon_2\approx U_2/\left(a g_0\right)\approx\Omega^2a^2/3(GM_a/a)$ (using Eq. \ref{shape} with neglecting $\phi_2$), it leads to a slight broadening of the equatorial belt where sub-inertial GIWs are propagative towards the surface.  

\subsection{Oscillation eigenmodes}
Studying oscillation eigenmodes induce the use of the quantization as derived in \S \ref{sec:SD}. As discussed in the previous section, it would be relevant to expand the eigenfrequencies and the corresponding spin parameters as a combination of their values in the spherical case and of a centrifugal correction:
\begin{eqnarray}
\omega_{\nu km}&=&\omega_{\nu km}^{\left(0\right)}+\omega_{\nu km}^{\left(1\right)},\\
\nu_{km}&=&2\Omega/\omega_{\nu km}=\nu_{km}^{\left(0\right)}+\nu_{km}^{\left(1\right)}
\end{eqnarray}
with $\nu_{km}^{\left(0\right)}=2\Omega/\omega_{\nu km}^{\left(0\right)}$ and $\nu_{km}^{\left(1\right)}=-\left(2\Omega/\omega_{\nu km}^{\left(0\right)}\right)\left(\omega_{\nu km}^{\left(1\right)}/\omega_{\nu km}^{\left(0\right)}\right)$. Such an expansion should then be introduced in the expression of the vertical wave number ($k_{V;\nu k m}$; Eq. \ref{Eq:dispers}) and of ${\mathcal L}_{\nu m}^{\left(1\right)}$ (Eq. \ref{eq:PLTO}) that would lead to complex and heavy implicit equations to solve. Therefore, we advocate to solve directly the generalised Laplace tidal equation (Eq. \ref{tidal}) and Eq. (\ref{eigenfrequencies}) when studying the case of oscillation eigenmodes.

\end{appendix}

\end{document}